\documentclass[12pt,titlepage]{article}
\renewcommand{\to}{\rightarrow}
\newcommand{\beq}{\begin{equation}}
\newcommand{\eeq}{\end{equation}}
\newcommand{\bea}{\begin{eqnarray}}
\newcommand{\eea}{\end{eqnarray}}
\def\e{{\rm e}}

\def\tr{{\rm tr}}
\begin{document}
\thispagestyle{empty}
\begin{titlepage}
\addtolength{\baselineskip}{.7mm}
\thispagestyle{empty}
\begin{flushright}
\end{flushright}
\vspace{10mm}
\begin{center}
{\large 
{\bf An introduction to symmetric spaces
}}\\[15mm]
{\bf 
Ulrika~Magnea 
} \\
\vspace{5mm}
{\it Department of Mathematics, University of Torino \\
Via Carlo Alberto 10, I-10125 Torino, Italy }\\
magnea@dm.unito.it
\\[6mm]
\vspace{13mm}
{\bf Abstract}\\[5mm]
\end{center}

Recently, the theory of symmetric spaces has come to play an increased
role in the physics of integrable systems and in quantum transport
problems. In addition, it provides a classification of random matrix
theories.  In this paper we give a self--contained introduction
to symmetric spaces and their main characteristics. We take an
algebraic approach; therefore it is not necessary to know almost
anything about differential geometry to be able to follow the outline.

\end{titlepage}
\newpage
\setcounter{footnote}{0}

\section{Introduction}
\label{sec-Intro}

Recently, the study of symmetric spaces has gained renewed interest in
physics. This is mainly due to two developments. The first of these
connects random matrix theories to symmetric spaces, and provides a
new classification of the former.  The second development started in
the eighties with the work of Olshanetsky and Perelomov
\cite{OlshPere}, who demonstrated the deep connection between some
quantum integrable systems and root systems of Lie algebras.

The connection that has emerged between random matrix theories and
symmetric spaces was mentioned already by Dyson \cite{DysonSS}.  The
integration manifolds defining the symmetry classes of the random
matrix ensembles are usually symmetric spaces with positive or
negative curvature.  In contrast, the integration manifold could also
be the {\it algebra} of matrices spanning a symmetric space of zero
curvature.  These issues will be discussed below and in more detail in a
forthcoming paper \cite{CasMag}.

Although Dyson was the first to recognize that the integration
manifolds in random matrix theory actually are symmetric spaces, the
subsequent emergence of new random matrix symmetry classes and their
classification in terms of Cartan's symmetric spaces is relatively
recent \cite{AltZ,Zirn,MCclass,TitBrou,Ivanov}.  Until recently,
random matrix ensembles used for physical applications were known to
correspond to ten of the eleven classes of symmetric spaces in
Cartan's classification. In a recent paper, Ivanov \cite{Ivanov} found
realizations of the remaining class, the algebra ${\bf SO(2n+1)}$ as
well as ${\bf SO(4n+2)}/{\bf U(2n+1)}$ in disordered vortices in
$p$--wave superconductors (this author splits the Cartan class DIII
into DIII-even and DIII-odd; in that case there are twelve Cartan
classes and these algebras correspond to B and DIII-odd). Thereby each
Cartan class is realized in some physical system.

In the early eighties, Olshanetsky and Perelomov showed that these
same symmetric spaces, through their root systems, are related to
integrable Calogero--Sutherland models. In \cite{OlshPere} the authors
demonstrated that it is the symmetry of the underlying root systems
that make these models integrable for certain values of the coupling
constants in the Calogero--Sutherland potential. These special values
are determined by the multiplicities of the various types of roots
(long, ordinary, and short) in the corresponding root system. This
deep and beautiful connection between Lie algebras and integrability
of many--particle systems is described in \cite{OlshPere}. In
addition, the same authors showed that the dynamics of these quantum
integrable systems is related to free diffusion on symmetric spaces.

The above scenario was applied to the physics of disordered wires in
\cite{MCDMPK}, where it was demonstrated that the symmetry class of
the random matrix ensemble used in modeling the transfer matrix of a
disordered wire determines a particular Calogero--Sutherland model,
through the above mentioned common connection to a particular
symmetric space.  The Calogero--Sutherland Hamiltonian so defined maps
onto the radial part of the Laplace--Beltrami operator on the
underlying symmetric space \cite{OlshPere}. This is the connection to
the dynamics of the quantum system mentioned in the preceding
paragraph.  The Laplace--Beltrami operator defines free diffusion on
the symmetric space. The equation of free diffusion becomes the
Dorokhov--Mello--Pereyra--Kumar (DMPK) equation, a differential
equation describing the evolution of the transmission eigenvalues of
the disordered wire with increasing length of the wire.  As a
consequence, it was demonstrated in \cite{MCDMPK,MCdis} that known
properties of the underlying symmetric space and of the integrable
Calogero--Sutherland model can be exploited in solving the DMPK
equation for a disordered wire.

In a forthcoming publication \cite{CasMag} we will discuss the present
status of the various applications of symmetric spaces to transport
problems and in the field of integrable systems, indicating some new
directions of research.  In addition, we will discuss the applications
of random matrix theory and the classification of the ensembles of
random matrix theory implied by the Cartan classification of symmetric
spaces.

The theory of symmetric spaces has a long history in mathematics.
Here we will give a brief, self--contained introduction to symmetric
spaces, listing in the process some references that are more easily
accessible to physicists than the standard reference, the book by
Helgason \cite{Helgason}. The reader is referred to this book for a
rigorous treatment. We will concentrate on the issues that will be of
relevance in our forthcoming paper \cite{CasMag}.  For physicists with
little background in differential geometry we recommend the book by
Gilmore \cite{Gilmore} (especially Chapter~9) for an introduction of
exceptional clarity.  Our treatment will be somewhat rigorous;
however, we skip proofs that can be found in the mathematical
literature and concentrate on simple examples that illustrate the
concepts presented.

In section~\ref{sec-Lie}, after reviewing the basics about Lie groups,
we will present some of the most important properties of root systems.
In section~\ref{sec-strSS} we define symmetric spaces and discuss
their main characteristics, defining involutive automorphisms,
spherical decomposition of the group elements, and the metric on the
Lie algebra.  We also discuss the algebraic structure of the coset
space.

In section~\ref{sec-realforms} we show how to obtain all the real
forms of a complex semisimple Lie algebra. The same techniques will
then be used to classify the real forms of the symmetric spaces in
section~\ref{sec-claSS}. In this section we also define the curvature
of a symmetric space, and discuss triplets of symmetric spaces with
postive, zero and negative curvature, all corresponding to the same
symmetric subgroup.  We will see why curved symmetric spaces arise
from semisimple groups, whereas the flat spaces are associated to
non--semisimple groups.  In addition, in section~\ref{sec-claSS} we
will define restricted root systems.  The restricted root systems are
associated to symmetric spaces, just like ordinary root systems are
associated to groups. As we will discuss in detail in \cite{CasMag},
they are key objects when considering the integrability of
Calogero--Sutherland models.

In the last section of the paper we will discuss Casimir and Laplace
operators on symmetric spaces and mention some known properties of the
eigenfunctions of the latter, so called zonal spherical functions.
These functions play a prominent role in many physical applications.
In every section we work out several simple examples that illustrate
the material presented.  This paper contains the basis for
understanding the developments to be discussed in more detail in
\cite{CasMag}.

\section{Lie groups and root spaces}
\label{sec-Lie}

In this introductory section we define the basic concepts relating to
Lie groups. We will build on the material presented here when we
discuss symmetric spaces in the next main section of the paper. The
reader with a solid background in group theory may want to skip most
or all of this section.

\subsection{Lie groups and manifolds}

A manifold can be thought of as the generalization of a surface, but
we do not in general consider it as embedded in a higher--dimensional
euclidean space.  A short introduction to differentiable manifolds can
be found in ref.~\cite{FosNigh}, and a more elaborate one in
refs.~\cite{Boothby} and \cite{3w} (Ch.~III). The points of an
$N$--dimensional manifold can be labelled by real coordinates
$(x^1,...,x^N)$. Suppose that we take an open set $U_\alpha $ of this
manifold, and we introduce local real coordinates on it. Let
$\psi_\alpha $ be the function that attaches $N$ real coordinates to
each point in the open set $U_\alpha $.  Suppose now that the manifold
is covered by overlapping open sets, with local coordinates attached
to each of them. If for each pair of open sets $U_\alpha $, $U_\beta
$, the fuction $\psi_\alpha \circ \psi_\beta^{-1}$ is differentiable
in the overlap region $U_\alpha \cap U_\beta $, it means that we can
go smoothly from one coordinate system to another in this region. Then
the manifold is differentiable.

Consider a group $G$ acting on a space $V$. We can think of $G$ as
being represented by matrices, and of $V$ as a space of vectors on
which these matrices act.  A group element $g\in G$ transforms the
vector $v\in V$ into $gv=v'$.  

If $G$ is a Lie group, it is also a differentiable manifold.  The fact
that a Lie group is a differentiable manifold means that for two group
elements $g$, $g'\in G$, the product $(g,g') \in G\times G \to gg'\in
G$ and the inverse $g\to g^{-1}$ are smooth ($C^\infty$) mappings,
that is, these mappings have continuous derivatives of all orders.

{\bf Example:} The space ${\bf R^n}$ is a smooth manifold and at the
same time an abelian group. The ``product'' of elements is addition
$(x,x')\to x+x'$ and the inverse of $x$ is $-x$. These operations are
smooth.  

{\bf Example:} 
The set $GL(n,R)$ of nonsingular real $n\times n$ matrices
$M$, ${\rm det}M\neq 0$, with matrix multiplication $(M,N)\to MN$ and
multiplicative matrix inverse $M\to M^{-1}$ is a non--abelian group
manifold. Any such matrix can be represented as $M=\e^{\sum_i t^iX_i}$
where $X_i$ are generators of the ${\bf GL(n,R)}$ algebra and $t^i$
are real parameters.

\subsection{The tangent space}

In each point of a differentiable manifold, we can define the tangent
space.  If a curve through a point $P$ in the manifold is parametrized
by $t\in {\bf R}$

\beq
x^a(t)=x^a(0)+\lambda^at \ \ \ \ \ \ \ \ a=1,...,N
\eeq

where $P=(x^1(0),...,x^N(0))$, then ${\bf
  \lambda}=(\lambda^1,...,\lambda^N)= (\dot{x}^1(0),...,\dot{x}^N(0))$
is a tangent vector at $P$. Here $\dot{x}^a(0) =\frac{\rm d}{{\rm
    d}t}x^a(t)|_{t=0}$.  The space spanned by all tangent vectors at
$P$ is the tangent space. In particular, the tangent vectors to the
coordinate curves (the curves obtained by keeping all the coordinates
fixed except one) through $P$ are called the natural basis for the
tangent space.  

{\bf Example:} In euclidean 3--space the natural basis is $\{ \hat
e_x,\hat e_y, \hat e_z\}$. On a patch of the unit 2--sphere parametrized
by polar coordinates it is $\{ \hat e_\theta,\hat e_\phi \}$.

For a Lie group, the tangent space at the origin is spanned by the
generators, that play the role of (contravariant) vector fields (also
called derivations), expressed in local coordinates on the group
manifold as $X=X^a(x)\partial_a$ (for an introduction to differential
geometry see ref.~\cite{SattW}, Ch. 5, or \cite{3w}). Here the partial
derivatives $\partial_a=\frac{\partial }{\partial x^a}$ form a basis
for the vector field.  That the generators span the tangent space at
the origin can easily be seen from the exponential map. Suppose $X$ is
a generator of a Lie group. The exponential map then maps $X$ onto
$\e^{tX}$, where $t$ is a parameter. This mapping is a one--parameter
subgroup, and it defines a curve $x(t)$ in the group manifold. The
tangent vector of this curve at the origin is then

\beq
\frac{\rm d}{{\rm d}t} \e^{tX}|_{t=0} =X
\eeq

All the generators together span the tangent space at the origin
(the identity element).

\subsection{Coset spaces}
\label{sec-cosets}

The isotropy subgroup $G_{v_0}$ of a group $G$ at the point $v_0\in V$
is the subset of group elements that leave $v_0$ fixed. The set of
points that can be reached by applying elements $g\in G$ to $v_0$ is
the orbit of $G$ at $v_0$, denoted $Gv_0$. If $Gv_0=V$ for one point
$v_0$, then this is true for every $v\in V$.  We then say that $G$
{\it acts transitively} on $V$.

In general, a symmetric space can be represented as a coset space.
Suppose $H$ is a subgroup of a Lie group $G$. The coset space $G/H$ is
the set of subsets of $G$ of the form $gH$, for $g\in G$. $G$ acts on
this coset space: $g_1(gH)$ is the coset $(g_1g)H$.

If $G$ acts transitively on $V$, then $V=Gv$ for any $v\in V$. Since
the isotropy subgroup $G_{v_0}$ leaves a fixed point $v_0$ invariant,
$gG_{v_0}v_0=gv_0=v\in V$, we see that the action of the group $G$ on
$V$ defines a bijective action of elements of $G/G_{v_0}$ on $V$.
Therefore the space $V$ on which $G$ acts transitively, can be
identified with $G/G_{v_0}$, since there is one--to--one
correspondence between the elements of $V$ and $G/G_{v_0}$. There is a
natural mapping from the group element $g$ onto the point $gv_0$ on
the manifold.

{\bf Example:} The $SO(2)$ subgroup of $SO(3)$ is the isotropy
subgroup at the north pole of a unit 2--sphere imbedded in
3--dimensional space, since it keeps the north pole fixed. On the
other hand, the north pole is mapped onto any point on the surface of
the sphere by elements of the coset $SO(3)/SO(2)$.  This can be seen
from the explicit form of the coset representatives.  As we will see
in eq.~(\ref{eq:cosetreps}) in subsection \ref{sec-algstr}, the
general form of the elements of the coset is

\beq
\label{eq:cosetrep}
M={\rm exp}\left(\begin{array}{cc} 0 & C \\ 
                          -C^T & 0 \end{array}\right)
=\left(\begin{array}{cc}\sqrt{I_2-XX^T} & X   \\ 
                                   -X^T &  \sqrt{1-X^TX} \end{array}\right)
\eeq

where $C$ is the matrix

\beq
C=\left(\begin{array}{c}t^2 \\ t^1\end{array}\right)
\eeq

and $t^1$, $t^2$ are real coordinates. $I_2$ in
eq.~(\ref{eq:cosetrep}) is the $2\times 2$ unit matrix.  For the coset
space $SO(3)/SO(2)$, $M$ is equal to

\beq
\label{eq:L1L2}
M={\rm exp}\left(\sum_{i=1}^2 t^iL_i\right), \ \ \ \ 
L_1= \frac{1}{2}\left(\begin{array}{ccc} 0&0&0\\ 0&0&1\\ 0&-1&0\end{array}\right), \ \ \ \ 
L_2=\frac{1}{2}\left(\begin{array}{ccc} 0&0&1\\ 0&0&0\\ -1&0&0\end{array}\right)
\eeq

The third $SO(3)$ generator

\beq
\label{eq:L3}
L_3=\frac{1}{2}\left(\begin{array}{ccc} 0&1&0\\ -1&0&0\\ 0&0&0\end{array}\right)
\eeq

spans the algebra of the stability subgroup $SO(2)$, that keeps the north pole 
fixed: 

\beq
\label{eq:fixedNP}
{\rm exp}(t^3L_3)\left(\begin{array}{c} 0 \\ 0 \\ 1 \end{array}\right)=
\left(\begin{array}{c} 0 \\ 0 \\ 1 \end{array}\right)
\eeq

The generators $L_i$ ($i=1,2,3$)
satisfy the $SO(3)$ commutation relations $[L_i,L_j]=\frac{1}{2}\epsilon_{ijk}L_k$. 
Note that since the $L_i$ and the $t^i$ are real, $C^\dagger =C^T$.

In (\ref{eq:cosetrep}), $M$ is a general representative of the coset
$SO(3)/SO(2)$.  By expanding the exponential we see that the explicit
form of $M$ is

\beq
M=\left(
  \begin{array}{ccc} 1+(t^2)^2\frac{({\rm cos}\sqrt{(t^1)^2+(t^2)^2}-1)}{(t^1)^2+(t^2)^2} &
                     t^1t^2\frac{({\rm cos}\sqrt{(t^1)^2+(t^2)^2}-1)}{(t^1)^2+(t^2)^2} &
                     t^2\frac{{\rm sin}\sqrt{(t^1)^2+(t^2)^2}}{\sqrt{(t^1)^2+(t^2)^2}}\\

                     t^1t^2\frac{({\rm cos}\sqrt{(t^1)^2+(t^2)^2}-1)}{(t^1)^2+(t^2)^2} &
                     1+(t^1)^2\frac{({\rm cos}\sqrt{(t^1)^2+(t^2)^2}-1)}{(t^1)^2+(t^2)^2} &
                     t^1\frac{{\rm sin}\sqrt{(t^1)^2+(t^2)^2}}{\sqrt{(t^1)^2+(t^2)^2}}\\

                     -t^2\frac{{\rm sin}\sqrt{(t^1)^2+(t^2)^2}}{\sqrt{(t^1)^2+(t^2)^2}} &
                     -t^1\frac{{\rm sin}\sqrt{(t^1)^2+(t^2)^2}}{\sqrt{(t^1)^2+(t^2)^2}} &
                     {\rm cos}\sqrt{(t^1)^2+(t^2)^2}\end{array}
\right)
\eeq

Thus the matrix $X=\left(\begin{array}{c}x\\ y\end{array}\right)$ is
given in terms of the components of $C$ by (cf.
eq.~(\ref{eq:functions})):

\beq 
X=\left(\begin{array}{c}x\\ y\end{array}\right)=\left(\begin{array}{c}
t^2\frac{{\rm sin}\sqrt{(t^1)^2+(t^2)^2}}{\sqrt{(t^1)^2+(t^2)^2}}\\
t^1\frac{{\rm sin}\sqrt{(t^1)^2+(t^2)^2}}{\sqrt{(t^1)^2+(t^2)^2}}\end{array}\right)
\eeq

Defining now $z={\rm cos}\sqrt{(t^1)^2+(t^2)^2}$, we see that 
the variables $x$, $y$, $z$ satisfy the equation of the 2--sphere:

\beq
x^2+y^2+z^2=1
\eeq

When the coset space representative $M$ acts on the north pole it is 
easily seen that the orbit is all of the 2--sphere:

\beq
M\left(\begin{array}{c} 0\\ 0\\ 1\end{array}\right) =
\left(\begin{array}{ccc}  .   &  .   & x \\
                          .   &  .   & y \\
                          .   &  .   & z \end{array}\right)
\left(\begin{array}{c} 0\\ 0\\ 1\end{array}\right) =
\left(\begin{array}{c} x\\ y\\ z\end{array}\right) 
\eeq

This shows that there is one--to--one correspondence between the
elements of the coset and the 2--sphere. The coset $SO(3)/SO(2)$ can
therefore be identified with a unit 2--sphere imbedded in
3--dimensional space.

\subsection{The Lie algebra and the adjoint representation}
\label{sec-Lie,adj}

A Lie algebra ${\bf G}$ is a vector space over a field $F$. 
Multiplication in the Lie algebra is given by the bracket
$[X,Y]$. It has the following properties:

\noindent [1] If $X$, $Y\in {\bf G}$, then $[X,Y]\in {\bf G}$,\\
\noindent [2] $[X,\alpha Y+\beta Z]=\alpha [X,Y]+\beta[X,Z]$ for $\alpha $,
$\beta \in F$, \\
\noindent [3] $[X,Y]=-[Y,X]$, \\
\noindent [4] $[X,[Y,Z]]+[Y,[Z,X]]+[Z,[X,Y]]=0$ (the Jacobi identity).

The algebra ${\bf G}$ generates a group through the exponential mapping.
A general group element is 

\beq
M={\rm exp}\left( \sum_it^iX_i\right);\ \ \ \ t^i\in F,\ X_i\in {\bf G}
\eeq
  
We define a mapping ${\rm ad} X$ from the Lie algebra to itself by
${\rm ad} X:Y\to [X,Y]$.  The mapping $X\to {\rm ad} X$ is a
representation of the Lie algebra called the adjoint representation.
It is easy to check that it is an automorphism: it follows from the
Jacobi identity that $[{\rm ad}X_i,{\rm ad}X_j]={\rm ad}[X_i,X_j]$.
Suppose we choose a basis $\{ X_i\}$ for ${\bf G}$. Then

\beq
\label{eq:adjr}
{\rm ad} X_i(X_j)=[X_i,X_j]=C^k_{ij}X_k
\eeq

where we sum over $k$. The $C^k_{ij}$ are called structure constants.
Under a change of basis, they transform as mixed tensor components.
They define the matrix $(M_i)_{jk}=C^j_{ik}$ associated with the
adjoint representation of $X_i$.  One can show that there exists a
basis for any complex semisimple algebra in which the structure
constants are real. This means the adjoint representation is real.
Note that the dimension of the adjoint representation is equal to the
dimension of the group.
 
{\bf Example:} Let's construct the adjoint representation of $SU(2)$.
The generators in the defining representation are

\beq
J_3=\frac{1}{2}\left( \begin{array}{cc} 1 & 0  \\ 
                                   0 & -1 \end{array} \right), \ \ \ \ \  
J_\pm =\frac{1}{2}\left( \left( \begin{array}{cc} 0 & 1  \\ 
                                   1 & 0  \end{array} \right)\pm 
i\left( \begin{array}{cc} 0 & -i \\ 
                                   i & 0  \end{array} \right)\right)
\eeq

and the commutation relations are

\beq
[J_3,J_\pm]=\pm J_\pm, \ \ \ \ \ \ \ [J_+,J_-]=2J_3
\eeq

The structure constants are therefore 
$C^+_{3+}=-C^+_{+3}=-C^-_{3-}=C^-_{-3}=1$, $C^3_{+-}=-C^3_{-+}=2$ and the
adjoint representation is given by $(M_3)_{++}=1$, $(M_3)_{--}=-1$, 
$(M_+)_{+3}=-1$, $(M_+)_{3-}=2$, $(M_-)_{-3}=1$, $(M_-)_{3+}=-2$, and all
other matrix elements equal to 0:

\beq
\label{eq:SU2adjoint}
M_3=\left(\begin{array}{ccc}0 & 0 & 0 \\0 & 1 & 0 \\ 0 & 0 & -1 \end{array}\right),
\ \ \ \ \ 
M_+=\left(\begin{array}{ccc}0 & 0 & 2 \\ -1 & 0 & 0\\ 0 & 0 & 0 \end{array}\right),
\ \ \ \ \ 
M_-=\left(\begin{array}{ccc} 0 & -2 & 0 \\ 0 & 0 & 0 \\ 1 & 0 & 0\end{array}\right),
\ \ \ \ \ 
\eeq

These representation matrices are real, have the same dimension as the
group, and satisfy the $SU(2)$ commutation relations $[M_3,M_\pm]=\pm
M_\pm $, $[M_+,M_-]=2M_3$.

\subsection{Semisimple algebras and root spaces}
\label{sec-rootsp}

In this paragraph we will briefly recall the basic facts about root
spaces and the classification of complex simple Lie algebras, to set
the stage for our discussion of real forms of Lie algebras and finally
symmetric spaces. 

An {\it ideal}, or {\it invariant subalgebra} ${\bf I}$ is a
subalgebra such that $[{\bf G},{\bf I}] \subset {\bf I}$.  An abelian
ideal also satisfies $[{\bf I},{\bf I}]=0$.  A {\it simple} Lie
algebra has no proper ideal. A {\it semisimple} Lie algebra is the
direct sum of simple algebras, and has no proper abelian ideal (by
proper we mean different from $\{ 0\}$).
  
A Lie algebra is a linear vector space over a field $F$, with an
antisymmetric product defined by the Lie bracket (cf. subsection
\ref{sec-Lie,adj}). If $F$ is the field of real, complex or quaternion
numbers, the Lie algebra is called a real, complex or quaternion
algebra.  A complexification of a real Lie algebra is obtained by
taking linear combinations of its elements with complex coefficients.
A real Lie algebra ${\bf H}$ is a real form of the complex algebra
${\bf G}$ if ${\bf G}$ is the complexification of ${\bf H}$.

In any simple algebra there are two kinds of generators: there is a
maximal abelian subalgebra, called the {\it Cartan subalgebra} ${\bf
  H_0}= \{ H_1,...,H_r\} $, $[H_i,H_j]=0$ for any two elements of the
Cartan subalgebra.  There are also raising and lowering operators denoted
$E_\alpha$. $\alpha $ is an $r$--dimensional vector $\alpha
=(\alpha_1,...,\alpha_r)$ and $r$ is the {\it rank} of the algebra.
\footnote{The rank of an algebra is defined through the secular
  equation (see subsection \ref{sec-Casi}).  For a non--semisimple
  algebra, the maximal number of mutually commuting generators can be
  greater than the rank of the algebra.}  The latter are
eigenoperators of the $H_i$ in the adjoint representation belonging to
eigenvalue $\alpha_i$: $[H_i,E_\alpha]=\alpha_iE_\alpha$.  For each
eigenvalue, or {\it root} $\alpha_i$, there is another eigenvalue
$-\alpha_i$ and a corresponding eigenoperator $E_{-\alpha}$ under the
action of $H_i$.

Suppose we represent each element of the Lie algebra
by an $n\times n$ matrix. Then $[H_i,H_j]=0$ means the matrices $H_i$
can all be diagonalized simultaneously. Their eigenvalues $\mu_i$ are
given by $H_i|\mu \rangle =\mu_i|\mu \rangle $, where the eigenvectors
are labelled by the {\it weight vectors} $\mu =(\mu_1,...,\mu_r)$
\cite{Georgi}.

A weight whose first non--zero component is positive is called a
positive weight.  Also, a weight $\mu $ is greater than another weight
$\mu' $ if $\mu - \mu' $ is positive. Thus we can define the highest
weight as the one which is greater than all the others. The highest
weight is unique in any representation.

The roots $\alpha_i \equiv \alpha(H_i)$ of the algebra ${\bf G}$
are the weights of the adjoint representation. Recall that in the
adjoint representation, the states on which the generators act are
defined by the generators themselves, and the action is defined by

\beq
X_a|X_b\rangle \equiv {\rm ad}X_a(X_b) \equiv [X_a,X_b]
\eeq

The roots are functionals on the Cartan subalgebra satisfying

\beq
{\rm ad}H_i(E_\alpha )=[H_i,E_\alpha]=\alpha (H_i)E_\alpha
\eeq

where $H_i$ is in the Cartan subalgebra.  The eigenvectors $E_\alpha$
are called the root vectors.  These are exactly the raising and
lowering operators $E_{\pm \alpha}$ for the weight vectors $\mu $.
There are canonical commutation relations defining the system of roots
belonging to each simple rank $r$--algebra. These are summarized
below: \footnote{For the reader who wants to understand more about the
  origin of the structure of Lie algebras, we recommend Chapter~7 of
  Gilmore \cite{Gilmore}.}

\beq
[H_i,H_j]=0,\ \ \ \ \ [H_i,E_\alpha]=\alpha_iE_\alpha, \ \ \ \ \ 
[E_\alpha,E_{-\alpha}]=\alpha_iH_i
\eeq

One can prove the fundamental relation \cite{SattW,Georgi}

\beq
\label{eq:fund}
\frac{2\alpha \cdot \mu}{\alpha^2}=-(p-q)
\eeq

where $\alpha $ is a root, $\mu $ is a weight, and $p$, $q$ are
positive integers such that $E_\alpha |\mu +p\alpha \rangle =0$,
$E_{-\alpha} |\mu -q\alpha \rangle =0$ \footnote{Here the scalar 
product $\cdot $ can be defined in terms of the metric on the Lie algebra.
For the adjoint representation, $\mu $ is a root $\beta $ and 

\beq
\frac{2\alpha \cdot \beta }{\alpha^2}=\frac{2K(H_\alpha, H_\beta )}
{K(H_\alpha, H_\alpha )}\equiv \frac{2 \beta (H_\alpha )}{\alpha  (H_\alpha )}
\eeq

\noindent where $K$ denotes the Killing form (see paragraph
\ref{sec-metric}).  There is always a unique element $H_\alpha $ in the
algebra such that $K(H,H_\alpha )=\alpha (H)$ for each $H\in {\bf H_0}$
(see for example \cite{SattW}, Ch.~10).  In general for a linear form
$\mu $ on the Lie algebra,

\beq
\frac{2\alpha \cdot \mu}{\alpha^2}=\frac{2\mu (H_\alpha)}
{\alpha (H_\alpha )}
\eeq

\noindent Then $\mu $ is a highest weight for some representation if and
only if this expression is an integer for each positive root $\alpha
$.}.  This relation gives rise to the strict properties of root
lattices, and permits the complete classification of all the complex
(semi)simple algebras.

Eq.~(\ref{eq:fund}) is true for any representation, but has
particularly strong implications for the adjoint representation. In
this case $\mu $ is a root.  As a consequence of eq.~(\ref{eq:fund}),
the possible angle between two root vectors of a simple Lie algebra is
limited to a few values: these turn out to be multiples of $\frac{\pi
  }{6}$ and $\frac{\pi }{4}$ (see e.g. \cite{Georgi}, Ch.~VI).  The
root lattice is invariant under reflections in the hyperplanes
orthogonal to the roots (the Weyl group).  As we will shortly see,
this is true not only for the root lattice, but for the weight lattice
of any representation.

Note that the roots $\alpha $ are real--valued linear functionals on
the Cartan subalgebra. Therefore they are in the space dual to ${\bf
  H_0}$.  A subset of the positive roots span the root lattice. These
are called simple roots. Obviously, since the roots are in the space
dual to ${\bf H_0}$, the number of simple roots is equal to the rank
of the algebra.

The same relation (\ref{eq:fund}) determines the highest weights of
all irreducible representations. Setting $p=0$, choosing a positive
integer $q$, and letting $\alpha $ run through the simple roots,
$\alpha=\alpha^i$ ($i=1,...,r$), we find the highest weights $\mu^i $
of all the irreducible representations corresponding to the given
value of $q$ \cite{Georgi}. For example, for $q=1$ we get the highest
weights of the $r$ fundamental representations of the group, each
corresponding to a simple root $\alpha^i$. For higher values of $q$ we
get the highest weights of higher--dimensional representations of the
same group.

The set of all possible simple root systems are classified by means of
Dynkin diagrams, each of which correspond to an equivalence class of
isomorphic Lie algebras.  The classical Lie algebras ${\bf
  SU(n+1,C)}$, ${\bf SO(2n+1,C)}$, ${\bf Sp(2n,C)}$ and ${\bf
  SO(2n,C)}$ correspond to root systems $A_n$, $B_n$, $C_n$, and
$D_n$, respectively.  In addition there are five exceptional algebras
corresponding to root systems $E_6$, $E_7$, $E_8$, $F_4$ and $G_2$.
Each of these complex algebras in general has several real forms
associated with it (see section \ref{sec-realforms}).  These real
forms correspond to the same Dynkin diagram and root system as the
complex algebra.  Since we will not make reference to Dynkin diagrams
in the following, we will not discuss them here.  The interested
reader can find sufficient material for example in the book by Georgi
\cite{Georgi}.

The (semi)simple complex algebra ${\bf G}$ decomposes into a direct
sum of root spaces \cite{SattW}:

\beq
\label{eq:diroot}
{\bf G}={\bf H_0}\oplus \sum_\alpha {\bf G_\alpha}
\eeq

where ${\bf G_\alpha}=\{ E_\alpha \}$, ${\bf G_{-\alpha }}=\{
E_{-\alpha }\}$. This will be evident in the example given below.

{\bf Example:} The root system $A_{n-1}$ corresponds to the complex
Lie algebra ${\bf SL(n,C)}$ and all its real forms. In a later section
we will see how to construct all the real forms associated with a
given complex Lie algebra. Let's see here explicitly how to construct
the root lattice of ${\bf SU(3,C)}$, which is one of the real forms of
${\bf SL(3,C)}$.

The generators are determined by the commutation relations. In physics
it is common to write the commutation relations in the form

\beq
[T_i,T_j]=if_{ijk}T_k
\eeq

(an alternative form is to define the generators as $X_i=iT_i$ and
write the commutation relations as $[X_i,X_j]=-f_{ijk}X_k$) where
$f_{ijk}$ are structure constants for the algebra ${\bf SU(3,C)}$.

Using the notation $g=\e^{it^aT_a}$ for the group elements (with $t^a$
real and a sum over $a$ implied), the generators $T_a$ in the
fundamental representation of this group are hermitean\footnote{Note
  that we have written an explicit factor of $i$ in front of the
  generators in the expression for the group elements. This is often
  done for compact groups; since the Killing form (subsection
  \ref{sec-metric}) has to be negative definite, the coordinates of
  the algebra spanned by the generators must be purely imaginary. Here
  we use this notation because it is conventional. If we absorb the
  factor of $i$ into the generators, we get {\it antihermitean}
  matrices $X_a=iT_a$; we will do this in the example in subsection
  \ref{sec-Inv} to comply with eq.~(\ref{eq:KP}). Of course, the
  matrices in the {\it algebra} are always antihermitean.}:
 
\beq
\label{eq:Gell-Mann}
\begin{array}{l}
T_1=\frac{1}{2}\left(\begin{array}{ccc} 0 & 1 & 0 \\
                                   1 & 0 & 0 \\
                                   0 & 0 & 0 \end{array}\right), \ \ \ \
T_2=\frac{1}{2}\left(\begin{array}{ccc} 0 & -i & 0 \\
                                   i & 0 & 0 \\
                                   0 & 0 & 0 \end{array}\right), \ \ \ \
T_3=\frac{1}{2}\left(\begin{array}{ccc} 1 & 0 & 0 \\
                                   0 & -1 & 0 \\
                                   0 & 0 & 0 \end{array}\right), \nonumber \\
\\
T_4=\frac{1}{2}\left(\begin{array}{ccc} 0 & 0 & 1 \\
                                   0 & 0 & 0 \\
                                   1 & 0 & 0 \end{array}\right), \ \ \ \  
T_5=\frac{1}{2}\left(\begin{array}{ccc} 0 & 0 & -i \\
                                   0 & 0 & 0 \\
                                   i & 0 & 0 \end{array}\right), \ \ \ \
T_6=\frac{1}{2}\left(\begin{array}{ccc} 0 & 0 & 0 \\
                                   0 & 0 & 1 \\
                                   0 & 1 & 0 \end{array}\right), \nonumber \\ 
\\ 
T_7=\frac{1}{2}\left(\begin{array}{ccc} 0 & 0 & 0 \\
                                   0 & 0 & -i \\
                                   0 & i & 0 \end{array}\right), \ \ \ \
T_8=\frac{1}{2\sqrt{3}}\left(\begin{array}{ccc} 1 & 0 & 0 \\
                                                0 & 1 & 0 \\
                                                0 & 0 & -2 
                                   \end{array}\right) \end{array}
\\ \nonumber
\eeq

In high energy physics the matrices $2T_a$ are known as Gell--Mann
matrices.  The generators are normalized in such a way that ${\rm
  tr}(T_aT_b)=\frac{1}{2} \delta_{ab}$. Note that $T_1$, $T_2$, $T_3$
form an ${\bf SU(2,C)}$ subalgebra.  We take the Cartan subalgebra to
be ${\bf H_0}=\{T_3,T_8\}$. The rank of this group is $r=2$.

Let's first find the weight vectors of the fundamental representation.
To this end we look for the eigenvalues $\mu_i $ of the operators in
the abelian subalgebra ${\bf H_0}$:

\beq 
T_3\left(\begin{array}{c} 1 \\ 0 \\ 0 \end{array}\right) = 
\frac{1}{2}\left(\begin{array}{c} 1 \\ 0 \\ 0 \end{array}\right),\ \ \ \ \ \ \
T_8\left(\begin{array}{c} 1 \\ 0 \\ 0 \end{array}\right) = 
\frac{1}{2\sqrt{3}}\left(\begin{array}{c} 1 \\ 0 \\ 0 \end{array}\right),
\eeq

therefore the eigenvector $(1\, 0\, 0)^T$ corresponds to the state
$|\mu \rangle $ where

\beq 
\mu \equiv (\mu_1,\mu_2)=\left(\frac{1}{2},\frac{1}{2\sqrt{3}}\right)
\eeq 

is distinguished by its eigenvalues under the operators $H_i$ of the
Cartan subalgebra. In the same way we find that $(0\, 1\, 0)^T$ and
$(0\, 0\, 1)^T$ correspond to the states labelled by weight vectors

\beq 
\mu' = \left(-\frac{1}{2},\frac{1}{2\sqrt{3}}\right) ,\ \ \ \ \ \ \ 
\mu''= \left(0,-\frac{1}{\sqrt{3}}\right)
\eeq

respectively. $\mu$, $\mu'$, and $\mu''$ are the weights of the
fundamental representation $\rho =D$ and they form an equilateral
triangle in the plane. The highest weight of the representation $D$ is
$\mu =\left(\frac{1}{2},\frac{1}{2\sqrt{3}}\right)$.

There is also another fundamental representation $\bar{D}$ of the
algebra ${\bf SU(3,C)}$, since it generates a group of rank 2.
Indeed, from eq.~(\ref{eq:fund}), for $p=0$, $q=1$, there is one
highest weight $\mu^i$, and one fundamental representation, for each
simple root $\alpha^i$.  The highest weight $\bar{\mu }$ of the
representation $\bar{D}$ is

\beq
\bar{\mu }= \left(\frac{1}{2},-\frac{1}{2\sqrt{3}}\right) 
\eeq

The highest weights of the representations corresponding to any
positive integer $q$ can be obtained as soon as we know the simple
roots.  Then, by operating with lowering operators on this weight, we
obtain other weights, on which we can further operate with lowering
operators until we have obtained all the weights in the
representation. For an example of this procedure see \cite{Georgi},
Ch.~IX.

Let's see now how to obtain the roots of ${\bf SU(3,C)}$. Each root
vector $E_\alpha $ corresponds to either a raising or a lowering
operator: $E_\alpha $ is the eigenvector belonging to the root
$\alpha_i \equiv \alpha (H_i)$ under the adjoint representation of
$H_i$, like in eq.~(\ref{eq:eig}).  Each raising or lowering operator
is a linear combination of generators $T_i$ that takes one state of
the fundamental representation to another state of the same
representation: $E_{\pm \alpha}|\mu \rangle = N_{\pm \alpha,\mu }|\mu
\pm \alpha \rangle $.  Therefore the root vectors $\alpha $ will be
differences of weight vectors in the fundamental representation.  We
find the raising and lowering operators $E_{\pm \alpha}$ to be

\beq
\label{eq:raislowSU3}
\begin{array}{l}
E_{\pm(1,0)}=\frac{1}{\sqrt{2}}(T_1\pm iT_2) \\ \\
E_{\pm(\frac{1}{2},\frac{\sqrt{3}}{2})}=\frac{1}{\sqrt{2}}(T_4\pm iT_5) \\ \\
E_{\pm(-\frac{1}{2},\frac{\sqrt{3}}{2})}=\frac{1}{\sqrt{2}}(T_6\pm iT_7) 
\end{array}
\\ \nonumber 
\eeq

These form the subspaces ${\bf G_\alpha}$ in eq.~(\ref{eq:diroot}).
In the fundamental representation, we find using the Gell--Mann
matrices that these are matrices with only one non--zero element. For
example, the raising operator $E_\alpha $ that corresponds to the root
$\alpha = (1,0)$ is

\beq
\label{eq:explE}
E_{+(1,0)}=\frac{1}{\sqrt{2}}\left(\begin{array}{ccc} 0 & 1 & 0 \\
                                                      0 & 0 & 0 \\
                                                      0 & 0 & 0 \end{array}\right)
\eeq

This operator takes us from the state $|\mu' \rangle =|-\frac{1}{2},
\frac{1}{2\sqrt{3}}\rangle$ to the state $|\mu \rangle =
|\frac{1}{2},\frac{1}{2\sqrt{3}}\rangle $.  The components of the root
vectors of ${\bf SU(3,C)}$ are the eigenvalues $\alpha_i$ of these
under the adjoint representation of the Cartan subalgebra. That is,

\beq
\label{eq:eig}
H_i|E_\alpha \rangle \equiv {\rm ad}H_i (E_\alpha )\equiv 
[H_i,E_\alpha] =\alpha_i |E_\alpha \rangle 
\eeq

This way we easily find the roots: we can either explicitly use the
structure constants of $SU(3)$ in
$[T_a,T_b]=if_{abc}T_c=-iC^c_{ab}T_c$ (note the explicit factor of $i$
due to our conventions regarding the generators) or we can use an
explicit representation for $H_i$, $E_\alpha $ like in
eqs.~(\ref{eq:Gell-Mann}), (\ref{eq:raislowSU3}), (\ref{eq:explE}), to
calculate the commutators:

\beq
\begin{array}{l}
{\rm ad}H_1(E_{\pm(1,0)})= [H_1,E_{\pm(1,0)}] = 
[T_3,\frac{1}{\sqrt{2}}(T_1\pm iT_2)] = \frac{1}{\sqrt{2}}(iT_2 \pm T_1)=\pm
E_{\pm(1,0)} \equiv \alpha^\pm_1 E_{\pm(1,0)} \\ \\
{\rm ad}H_2(E_{\pm(1,0)})= [H_2,E_{\pm(1,0)}] = 
[T_8,\frac{1}{\sqrt{2}}(T_1\pm iT_2)] = 0  \equiv \alpha^\pm_2 E_{\pm(1,0)}
\end{array} 
\eeq

The root vector corresponding to the raising operator $E_{+(1,0)}$ is
thus $\alpha =(\alpha^+_1,\alpha^+_2) =(1,0)$ and the root vector
corresponding to the lowering operator $E_{-(1,0)}$ is $-\alpha
=(\alpha^-_1,\alpha^-_2)=(-1,0)$. These root vectors are indeed the
differences between the weight vectors $\mu =
\left(\frac{1}{2},\frac{1}{2\sqrt{3}}\right)$ and $\mu' =
\left(-\frac{1}{2},\frac{1}{2\sqrt{3}}\right)$ of the fundamental
representation.

In the same way we find the other root vectors $\left(\pm\frac{1}{2},
  \pm\frac{\sqrt{3}}{2}\right)$,
$\left(\mp\frac{1}{2},\pm\frac{\sqrt{3}}{2}\right)$, and $(0,0)$ (with
multiplicity 2), by operating with $H_1$ and $H_2$ on the remaining
$E_{\pm \alpha }$'s and on the $H_i$'s.  The last root with
multiplicity 2 has as its components the eigenvalues under $H_1$,
$H_2$ of the states $|H_1\rangle $ and $|H_2\rangle $: $H_i|H_j\rangle
= [H_i,H_j]=0$; $i$, $j\in \{1,2\}$.  The root vectors form a regular
hexagon in the plane. The positive roots are $(1,0)$,
$\alpha^1=\left(\frac{1}{2},\frac{\sqrt{3}}{2}\right)$ and
$\alpha^2=\left(\frac{1}{2},-\frac{\sqrt{3}}{2}\right)$.  The latter
two are simple roots.  $(1,0)$ is not simple because it is the sum of
the other positive roots. There are two simple roots, since the rank
of $SU(3)$ is 2 and the root lattice is two--dimensional.

The root lattice of $SU(3)$ is invariant under reflections in the
hyperplanes orthogonal to the root vectors. This is true of any weight
or root lattice; the symmetry group of reflections in hyperplanes
orthogonal to the roots is called the {\it Weyl group}.  It is
obtained from eq.~(\ref{eq:fund}): since for any root $\alpha $ and
any weight $\mu $, $2(\alpha \cdot \mu )/\alpha^2$ is the integer
$q-p$,

\beq
\label{eq:Weyl}
\mu'=\mu - \frac{2(\alpha \cdot \mu )}{\alpha^2}\alpha
\eeq

is also a weight. Eq.~(\ref{eq:Weyl}) is exactly the above mentioned
reflection, as can easily be seen.

We have just shown by an example how to obtain a root system of type
$A_n$.  In general for any simple algebra the commutation relations
determine the Cartan subalgebra and raising and lowering operators,
that in turn determine a unique root system, and correspond to a given
Dynkin diagram. In this way we can classify all the simple algebras
according to the type of root system it possesses. The root systems
for the four infinite series of classical non--exceptional Lie groups
can be characterized as follows \cite{Georgi} (denote the
$r$--dimensional space spanned by the roots by ${\cal V}$ and let
$\{e_1,...e_n\}$ be a canonical basis in ${\bf R}^n$):

$A_{n-1}$: Let ${\cal V}$ be the hyperplane in ${\bf R}^n$ that passes
through the points $(1,0,0,...0)$, $(0,1,0,...,0)$, ...,
$(0,0,...,0,1)$ (the endpoints of the $e_i$, $i=1,...,n$). Then the
root lattice contains the vectors $\{ e_i-e_j, i\neq j \}$.

$B_n$:  Let ${\cal V}$ be ${\bf R}^n$; then the roots are $\{ \pm e_i,
\pm e_i \pm e_j, i\neq j \}$.

$C_n$: Let ${\cal V}$ be ${\bf R}^n$; then the roots are $\{ \pm 2e_i,
\pm e_i \pm e_j, i\neq j \}$.

$D_n$: Let ${\cal V}$ be ${\bf R}^n$; then the roots are $\{
\pm e_i \pm e_j, i\neq j \}$.

The root lattice $BC_n$, that we will discuss in conjunction with restricted
root systems, is the union of $B_n$ and $C_n$. It is characterized as follows:

$BC_n$:  Let ${\cal V}$ be ${\bf R}^n$; then the roots are $\{ \pm e_i, 
\pm 2e_i, \pm e_i \pm e_j, i\neq j \}$. 

Because this system contains both $e_i$ and $2e_i$, it is called
non--reduced (normally the only root collinear with $\alpha$ is
$-\alpha$).  However, it is irreducible in the usual sense, which
means it is not the direct sum of two disjoint root systems $B_n$ and
$C_n$.  This can be seen from the root multiplicities (cf. Table~1).

The semisimple algebras are direct sums of simple ones. That means the
simple constituent algebras commute with each other, and the root
systems are direct sums of the corresponding simple root systems.
Therefore, knowing the properties of the simple Lie algebras, we also
know the semisimple ones.

\section{Symmetric spaces}
\label{sec-strSS}

In the previous section, we have reminded ourselves of some elementary
facts concerning root spaces and the classification of the complex semisimple
algebras. In this section we will define and discuss symmetric spaces.

A symmetric space is associated to an involutive automorphism of a
given Lie algebra. As we will see, several different involutive
automorphisms can act on the same algebra. Therefore we normally have
several different symmetric spaces deriving from the same Lie algebra.
The involutive automorphism defines a symmetric subalgebra and a
remaining complementary subspace of the algebra.  Under general
conditions, the complementary subspace is mapped onto a symmetric
space through the exponential map. In the following subsections we
make these statements more precise. We discuss how the elements of the
Lie group can act as transformations on the elements of the symmetric
space. This naturally leads to the definition of two coordinate
systems on symmetric spaces: the spherical and the horospheric
coordinate systems. The radial coordinates associated to each element
on a symmetric space through their spherical or horospheric
decomposition will be of relevance when we discuss the radial parts of
differential operators on symmetric spaces in section
\ref{sec-Operators}. In the same section we explain why these
operators are important in applications to physical problems, and in
\cite{CasMag} we will discuss some of their uses.

At the end of this section we define the metric tensor on a Lie
algebra in terms of the Killing form. The latter is defined as a
symmetric bilinear trace form on the adjoint representation, and is
therefore expressible in terms of the structure constants. We will
give several examples of Killing forms later, as we discuss the
various real forms of a Lie algebra. The metric tensor will serve to
define the curvature tensor on a symmetric space (subsection
\ref{sec-curv}). It is also needed in computing the Jacobian of the
transformation to radial coordinates. This Jacobian is relevant in
calculating the radial part of the Laplace--Beltrami operator (see
paragraph \ref{sec-Laplaceop}). Finally we discuss the general
algebraic form of coset representatives in subsection \ref{sec-algstr}.

\subsection{Involutive automorphisms}
\label{sec-Inv}

An automorphism of a Lie algebra ${\bf G}$ is a mapping from ${\bf G}$
onto itself such that it preserves the algebraic operations on the Lie
algebra. For example, if $\sigma $ is an automorphism, it preserves
multiplication: $[\sigma (X), \sigma (Y)]=\sigma ([X,Y])$, for $X$,
$Y\in {\bf G}$.

Suppose that the linear automorphism $\sigma:{\bf G}\to{\bf G}$ is
such that $\sigma^2=1$, but $\sigma $ is not the identity. That means
that $\sigma $ has eigenvalues $\pm 1$, and it splits the algebra
${\bf G}$ into orthogonal eigensubspaces corresponding to these
eigenvalues.  Such a mapping is called an {\it involutive
  automorphism}.

Suppose now that ${\bf G}$ is a compact simple Lie algebra, $\sigma $
is an involutive automorphism of ${\bf G}$, and ${\bf G}={\bf K}
\oplus {\bf P}$ where

\beq
\label{eq:KP}
\sigma (X)=X\ \  {\rm for}\ \  X\in {\bf K},\ \ \sigma (X)=-X\ \  
{\rm for}\ \  X\in {\bf P}
\eeq

From the properties of automorphisms mentioned above, it is easy to
see that ${\bf K}$ is a subalgebra, but ${\bf P}$ is not.  In fact,
the commutation relations

\beq
\label{eq:commrel}
[{\bf K},{\bf K}]\subset {\bf K},\ \  [{\bf K},{\bf P}]\subset {\bf P},\ \  
[{\bf P},{\bf P}]\subset {\bf K} 
\eeq

hold. A subalgebra ${\bf K}$ satisfying (\ref{eq:commrel}) is called a
{\it symmetric subalgebra}.  If we now multiply the elements in ${\bf
  P}$ by $i$ (the ``Weyl unitary trick''), we construct a new
noncompact algebra ${\bf G^*}={\bf K} \oplus i{\bf P}$. This is called
a {\it Cartan decomposition}, and ${\bf K}$ is a maximal compact
subalgebra of ${\bf G^*}$.  The coset spaces $G/K$ and $G^*/K$ are
{\it symmetric spaces}.

{\bf Example}: Suppose $G=SU(n,C)$, the group of unitary complex
matrices with determinant $+1$. The algebra of this group then
consists of complex antihermitean\footnote{See the footnote in
  subsection \ref{sec-rootsp}.}  matrices of zero trace (this follows
by differentiating the identities $UU^\dagger=1$ and ${\rm det}U=1$
with respect to $t$ where $U(t)$ is a curve passing through the
identity at $t=0$); a group element is written as $g=\e^{t^aX_a}$ with
$t^a$ real.  Therefore any matrix $X$ in the Lie algebra of this group
can be written $X=A+iB$, where $A$ is real, skew--symmetric, and
traceless and $B$ is real, symmetric and traceless.  This means the
algebra can be decomposed as ${\bf G}={\bf K}\oplus {\bf P}$, where
${\bf K}$ is the compact connected subalgebra ${\bf SO(n,R)}$
consisting of real, skew--symmetric and traceless matrices, and ${\bf
  P}$ is the subspace of matrices of the form $iB$, where $B$ is real,
symmetric, and traceless. ${\bf P}$ is not a subalgebra.

Referring to the example for ${\bf SU(3,C)}$ in subsection
\ref{sec-rootsp} we see, setting $X_a = iT_a$, that the $\{ X_a\} $
split into two sets under the involutive automorphism $\sigma $
defined by complex conjugation $\sigma =K$. This splits the compact
algebra ${\bf G}$ into ${\bf K}\oplus {\bf P}$, since ${\bf P}$
consists of imaginary matrices:

\beq
\label{eq:SU3}
\begin{array}{l}
{\bf K}=\{ X_2,X_5,X_7\}
=\left\{
\frac{1}{2}\left(\begin{array}{ccc} 0&1&0\\ -1&0&0\\ 0&0&0\end{array}\right),
\frac{1}{2}\left(\begin{array}{ccc} 0&0&1\\ 0&0&0\\ -1&0&0\end{array}\right),
\frac{1}{2}\left(\begin{array}{ccc} 0&0&0\\ 0&0&1\\ 0&-1&0\end{array}\right)
\right\}\\
\\
{\bf P}=\{ X_1,X_3,X_4,X_6,X_8\}\\ \\
=\left\{
\frac{i}{2}\left(\begin{array}{ccc} 0&1&0\\1&0&0\\0&0&0\end{array}\right),
\frac{i}{2}\left(\begin{array}{ccc} 1&0&0\\0&-1&0\\0&0&0\end{array}\right),
\frac{i}{2}\left(\begin{array}{ccc} 0&0&1\\0&0&0\\1&0&0\end{array}\right),
\frac{i}{2}\left(\begin{array}{ccc} 0&0&0\\0&0&1\\0&1&0\end{array}\right),
\frac{i}{2\sqrt{3}}\left(\begin{array}{ccc} 1&0&0\\0&1&0\\0&0&-2\end{array}\right)
\right\} \end{array} \nonumber \\ 
\nonumber \\ 
\eeq

${\bf K}$ spans the real subalgebra ${\bf SO(3,R)}$.  Setting $X_2
\equiv L_3$, $X_5 \equiv L_2$, $X_7 \equiv L_1$, the commutation
relations for the subalgebra are
$[L_i,L_j]=\frac{1}{2}\epsilon_{ijk}L_k$.  The Cartan subalgebra
$i{\bf H_0}=\{ X_3,X_8\} $ is here entirely in the subspace ${\bf P}$.

Going back to the general case of ${\bf G}={\bf SU(n,C)}$, we obtain
from ${\bf G}$ by the Weyl unitary trick the non--compact algebra
${\bf G^*}={\bf K}\oplus i{\bf P}$. $i{\bf P}$ is now the subspace of
real, symmetric, and traceless matrices $B$. The Lie algebra ${\bf
  G^*}={\bf SL(n,R)}$ is then the set of $n\times n$ real matrices of
zero trace, and generates the linear group of transformations
represented by real $n\times n$ matrices of unit determinant.

The involutive automorphism that split the algebra ${\bf G}$ above was
defined to be complex conjugation $\sigma =K$.  The involutive
automorphism that splits ${\bf G^*}$ is defined by $\tilde{\sigma
  }(g)=(g^T)^{-1}$ for $g\in G^*$, as we will now see.  On the level
of the algebra, $\tilde{\sigma }(g)=(g^T)^{-1}$ means $\tilde{\sigma
  }(X)=-X^T$. Suppose now $g=\e^{tX}\in G^*$ with $X$ real and
traceless and $t$ a real parameter.  If now $X$ is an element of the
subalgebra ${\bf K}$, we then have $\tilde{\sigma }(X)=+X$, i.e.
$-X^T=X$ and $X$ is skew--symmetric.  If instead $X\in i{\bf P}$, we
have $\tilde{\sigma }(X)=-X^T=-X$, i.e. $X$ is symmetric. The
decomposition ${\bf G^*} = {\bf K} \oplus i{\bf P}$ is the usual
decomposition of a ${\bf SL(n,R)}$ matrix in symmetric and
skew--symmetric parts.

$G/K=SU(n,C)/SO(n,R)$ is a symmetric space of compact type, and the
related symmetric space of non--compact type is
$G^*/K=SL(n,R)/SO(n,R)$.

\subsection{The action of the group on the symmetric space}
\label{sec-action}

Let $G$ be a semisimple Lie group and $K$ a compact symmetric
subgroup.  As we saw in the preceding paragraph, the coset spaces
$G/K$ and $G^*/K$ represent symmetric spaces. Just as we have defined
a Cartan subalgebra and the rank of a Lie algebra, we can define, in
an exactly analogous way, a Cartan subalgebra and the rank of a
symmetric space.  A Cartan subalgebra of a symmetric space is a
maximal abelian subalgebra of the subspace ${\bf P}$ (see paragraph
\ref{sec-restricted}), and the rank of a symmetric space is the number
of generators in this subalgebra.

If $G$ is connected and ${\bf G} = {\bf K} \oplus {\bf P}$ where ${\bf
  K}$ is a compact symmetric subalgebra, then each group element can
be decomposed as $g=kp$ (right coset decomposition) or $g=pk$ (left
coset decomposition), with $k\in K={\rm e}^{\bf K}$, $p\in P={\rm
  e}^{\bf P}$.  $P$ is not a subgroup, unless it is abelian and
coincides with its Cartan subalgebra. However, if the involutive
automorphism that splits the algebra is denoted $\sigma $, one can
show (\cite{Hermann}, Ch. 6) that $gp\sigma (g^{-1})\in P$.  This
defines $G$ as a transformation group on $P$.  Since $\sigma
(k^{-1})=k^{-1}$ for $k\in K$, this means

\beq
p'=kpk^{-1}\in P 
\eeq

if $k\in K$, $p\in P$. Now suppose there are no other elements in $G$
that satisfy $\sigma (g)=g$ than those in $K$.  This will happen if
the set of elements satisfying $\sigma (g)=g$ is connected. Then $P$
is isomorphic to $G/K$. Also, $G$ acts transitively on $P$ in the
manner defined above (cf. subsection \ref{sec-cosets}).  The tangent
space of $G/K$ at the origin (identity element) is spanned by the
subspace ${\bf P}$ of the algebra.

\subsection{Radial coordinates}
\label{sec-radial}

In this paragraph we define two coordinate systems frequently used on
symmetric spaces.  Let ${\bf G}={\bf K}\oplus {\bf P}$ be a Cartan
decomposition of a semisimple algebra and let ${\bf H_0}\subset {\bf
  P}$ be a maximal abelian subalgebra in the subspace ${\bf P}$.
Define $M$ to be the subgroup of elements in $K$ such that

\beq
M=\{ k\in K:kHk^{-1}=H,\ H\in {\bf H_0}\}
\eeq

This set is called the centralizer of ${\bf H_0}$ in $K$. Under
conjugation by $k\in K$, each element $H$ of the Cartan subalgebra is
preserved. Further, denote

\beq
M'=\{ k\in K:kHk^{-1}=H',\ H,\, H'\in {\bf H_0}\}
\eeq

This is a larger subgroup than $M$ that preserves the Cartan
subalgebra as a whole, but not necessarily each element separately,
and is called the normalizer of ${\bf H_0}$ in $K$.  If $K$ is a
compact symmetric subgroup of $G$, one can show (\cite{Hermann},
Ch.~6) that every element $p$ of $P\simeq G/K$ is conjugated with some
element $h={\rm e}^H$ for some $H\in {\bf H_0}$ by means of the
adjoint representation\footnote{Note that \beq {\rm e}^{K}H{\rm
    e}^{-K}={\rm e}^{{\rm ad}K}H\equiv \sum_{n=0}^\infty \frac{({\rm
      ad}K)^n}{n!}H \eeq } of the stationary subgroup $K$:

\beq
\label{eq:KAK}
p=khk^{-1} = kh\sigma (k^{-1})
\eeq

where $k\in K/M$ and $H$ is defined up to the elements in the factor
group $M'/M$. This factor group coincides with the Weyl group that was
defined in eq.~(\ref{eq:Weyl}). It transforms an element of the algebra
${\bf H_0}$ into another element of the same algebra.  In fact, this means 
that every element $g\in G$ can be decomposed as $g=pk=k'hk'^{-1}k=k'hk''$.
This is very much like the Euler angle decomposition of $SO(n)$.  

Thus, if $x_0$ is the fixed point of the subgroup $K$, an arbitrary
point $x\in P$ can be written

\beq
x=khk^{-1}x_0=khx_0
\eeq

The coordinates $(k(x),h(x))$ are called spherical coordinates. $k(x)$
is the angular coordinate and $h(x)$ is the {\it spherical radial
  coordinate} of the point $x$.  Eq.~(\ref{eq:KAK}) defines the so
called spherical decomposition of the elements in the coset space.  Of
course, a similar reasoning is true for the space $P^*\simeq G^*/K$.

This means every matrix $p$ in the coset space $G/K$ can be {\it
  diagonalized} by a similarity transformation by the subgroup $K$,
and the radial coordinates are exactly the set of eigenvalues of the
matrix $p$.  These ``eigenvalues'' are not necessarily real numbers.
This is easily seen in the example in eq.~(\ref{eq:SU3}). It can also
be seen in the adjoint representation. Suppose the algebra ${\bf
  G}={\bf K}\oplus {\bf P}$ is compact.  From eq.~(\ref{eq:adjr}), in
the adjoint representation $H_i\in {\bf H_0}$ has the form

\beq
\label{eq:adjHi}
H_i=\left(\begin{array}{cccccccc} 0 & ...    &   & & & & & \\
                                . & \ddots &   & & & & & \\
                                . &        & 0 & & & & & \\
                                  &        &   & \alpha_i & & & & \\
                                  &        &   &          & -\alpha_i & & \\
                                  &        &   &          &           & \ddots & \\
                                  &        &   &          &           &        &  \eta_i & \\
                                  & & & & & &                                            & -\eta_i 
\end{array}\right)
\eeq

where the matrix is determined by the structure constants
($[H_i,H_j]=0$, $[H_i,E_{\pm \alpha }]=\pm \alpha_i E_{\pm \alpha}$
... and $\pm \alpha_i,...,\pm \eta_i$ are the roots corresponding to
$H_i$). Since the Killing form must be negative (see subsection
\ref{sec-metric}) for a compact algebra, the coordinates of the Cartan
subalgebra must be purely imaginary and the group elements
corresponding to ${\bf H_0}$ must have the form

\beq
\label{eq:adjrad}
\e^{i{\bf t\cdot H}}=\left(\begin{array}{cccccc} 1 & ...    &   & & & \\               
                   . & \ddots &   & & & \\
                   . &        & 1 & & & \\
                     &        &   & \e^{i{\bf t\cdot \alpha }} & & \\
                     &        &   &          & \ddots & \\
                     &        &   &          &        & 
\e^{-i{\bf t\cdot \eta }} \end{array}\right)
\eeq

with ${\bf t}=(t^1,t^2,...t^r)$ and $t^i$ real parameters.  In
particular, if the eigenvalues are real for $p \in P^*$, they are
complex numbers for $p\in P$.

{\bf Example}: In the example we gave in the preceding subsection, the
coset space $G^*/K$ $=SL(n,R)/SO(n)$ $\simeq P^*=\e^{i{\bf P}}$
consists of real positive--definite symmetric matrices.  Note that
${\bf G} = {\bf K} \oplus {\bf P}$ implies that $G$ can be decomposed
as $G=PK$ and $G^*$ as $G^*=P^*K$.  The decomposition $G^*=P^*K$ in
this case is the decomposition of a $SL(n,R)$ matrix in a
positive--definite symmetric matrix and an orthogonal one. Each
positive--definite symmetric matrix can be further decomposed: it can
be diagonalized by an $SO(n)$ similarity transformation. This is the
content of eq.~(\ref{eq:KAK}) for this case, and we know it to be true
from linear algebra.  Similarly, according to eq.~(\ref{eq:KAK}) the
complex symmetric matrices in $G/K$ $=SU(n,C)/SO(n)$ $\simeq P=\e^{\bf
  P}$ can be diagonalized by the group $K=SO(n)$ to a form where the
eigenvalues are similar to those in eq.~(\ref{eq:adjrad}).

In terms of the subspace ${\bf P}$ of the algebra, eq.~(\ref{eq:KAK})
amounts to saying that any two Cartan subalgebras ${\bf H_0}$, ${\bf
  H'_0}$ of the symmetric space are conjugate under a similarity
transformation by $K$, and we can choose the Cartan subalgebra in any
way we please.  However, the number of elements that we can
diagonalize simultaneously will always be equal to the rank of the
symmetric space.

There is also another coordinate system valid only for spaces of the
type $P^*\sim G^*/K$. This coordinate system is called {\it
  horospheric} and is based on the so called {\it Iwasawa
  decomposition} of the algebra:

\beq
{\bf G}={\bf N^+}\oplus {\bf H_0} \oplus {\bf K}
\eeq

Here ${\bf K},\ {\bf H_0},\ {\bf N^+}$ are three subalgebras of ${\bf
  G}$. ${\bf K}$ is a maximal compact subalgebra, ${\bf H_0}$ is a
Cartan subalgebra, and

\beq
{\bf N^+}=\sum_{\alpha \in R^+}{\bf G}_\alpha
\eeq

is an algebra of raising operators corresponding to the positive roots
$\alpha (H) >0$ with respect to ${\bf H_0}$. As a consequence, the
group elements can be decomposed $g=nhk$, in an obvious
notation. This means that if $x_0$ is the fixed point of $K$, any
point $x\in G^*/K$ can be written
 
\beq
x=nhkx_0=nhx_0
\eeq

The coordinates $(n(x),h(x))$ are called horospheric
coordinates and the element $h=h(x)$ is called the
{\it horospheric projection} of the point $x$ or the {\it horospheric
  radial coordinate}.

\subsection{The metric on a Lie algebra} 
\label{sec-metric}
 
A metric tensor can be defined on a Lie algebra
\cite{Helgason,Gilmore, SattW,Hermann}. If $\{ X_i\}$ form a basis for
the Lie algebra ${\bf G}$, it is defined by

\beq
\label{eq:metric}
g_{ij}=K(X_i,X_j)\equiv \tr ({\rm ad}X_i {\rm ad}X_j)=C^r_{is}C^s_{jr}
\eeq

The symmetric bilinear form $K(X_i,X_j)$ is called the {\it Killing form}.  
It is intrinsically associated with the Lie algebra, and
since the Lie bracket is invariant under automorphisms of the algebra, 
so is the Killing form.

{\bf Example:} The generators $X_7 \equiv L_1$, $X_5 \equiv L_2$, $X_2
\equiv L_3$ of $SO(3)$ given in eq.~(\ref{eq:SU3}) obey the
commutation relations
$[L_i,L_j]=C_{ij}^kL_k=\frac{1}{2}\epsilon_{ijk}L_k$. From
eq.~(\ref{eq:metric}), the metric for this algebra is
$g_{ij}=-\frac{1}{2}\delta_{ij}$. The generators and the structure
constants can be normalized so that the metric takes the canonical
form $g_{ij}=-\delta_{ij}$.

Just like we defined the Killing form $K(X_i,X_j)$ for the algebra
${\bf G}$ in eq.~(\ref{eq:metric}) using the adjoint representation,
we can define a similar trace form $K_\rho$ and a metric tensor
$g_\rho $ for {\it any} representation $\rho $ by

\beq
\label{eq:g_rho}
g_{\rho,ij}=K_\rho (X_i,X_j) ={\rm tr}(\rho (X_i)\rho (X_j))
\eeq

where $\rho (X)$ is the matrix representative of the Lie algebra
element $X$.  If $\rho $ is an automorphism of ${\bf G}$, $K_\rho
(X_i,X_j)=K(X_i,X_j)$.

Suppose the Lie algebra is semisimple (this is true for all the
classical Lie algebras except the Lie algebras $GL(n,C)$, $U(n,C)$).
According to Cartan's criterion, {\it the Killing form is
  non--degenerate for a semisimple algebra}.  This means that ${\rm
  det}g_{ij}\neq 0$, so that the inverse of $g_{ij}$, denoted by
$g^{ij}$, exists. Since it is also real and symmetric, it can be
reduced to canonical form $g_{ij}={\rm diag}(-1,...,-1,1,...,1)$
with $p$ $-1$'s and $(n-p)$ $+1$'s, where $n$ is the dimension of the algebra. 

$p$ is an invariant of the quadratic form. In fact, for any real form
of a complex algebra, the trace of the metric, called the {\it
  character} of the particular real form (see below and in
\cite{Gilmore}) distinguishes the real forms from each other (though
it can be degenerate for the classical Lie algebras \cite{Gilmore}).
The character ranges from $-n$, where $n$ is the dimension of the
algebra, to $+r$, where $r$ is its rank. All the real forms of the
algebra have a character that lies in between these values. In subsection
\ref{sec-realforms1} we will see several explicit examples of Killing
forms.

A famous theorem by Weyl states that {\it a simple Lie group $G$ is
  compact, if and only if the Killing form on ${\bf G}$ is negative
  definite}. Otherwise it is non--compact. This is actually quite
intuitive and natural (see \cite{Gilmore}, Ch.~9, paragraph~I.2). On a
compact algebra, the metric can be chosen to be minus the Killing
form, if it is required to be positive--definite.

The metric on the Lie algebra can be extended to the whole coset space
$P\simeq G/K$, $P^*\simeq G^*/K$ as follows. At the origin of $G/K$
and $G^*/K$, the identity element $I$, the metric is identified with
the metric in the algebra, restricted to the respective tangent spaces
${\bf P}$, $i{\bf P}$.  Since the group acts transitively on the coset
space (cf. paragraph \ref{sec-cosets}), and the orbit of the origin is
the entire space, we can use a group transformation to map the metric
at the origin to any point $M$ in the space.  The metric tensor at $M$
will depend on the coset representative $M$.  It is given by

\beq
\label{eq:g(M)}
g_{rs}(M)=
g_{ij}(I)\frac{\partial x^i(I)}{\partial x^r(M)}\frac{\partial x^j(I)}{\partial x^s(M)}
\eeq

where $g_{ij}(I)$ is the metric at the origin (identity element) of
the coset space. (\ref{eq:g(M)}) follows from the invariance of the
line element $ds^2=g_{ij}dx^idx^j$ under translations.  If $\{ X_i\} $
is a basis in the tangent space, and $dM={\rm exp}(dx^iX_i)$ is a
coset representative infinitesimally close to the identity, we need to
know how $dx^i$ transforms under translations by the coset
representative $M$. We will not discuss that here, but some
generalities can be found for example in Ch.~9, paragraph V.4. of
ref.~\cite{Gilmore}. In general, it is not an easy problem unless the
coset has rank $1$.

{\bf Example:} The line element $ds^2$ on the radius--1 2--sphere
$SO(3)/SO(2)$ in polar coordinates is $ds^2= d\theta^2+{\rm
  sin}^2\theta \, d\phi^2$. The metric at the point $(\theta,\phi)$ is

\beq
\label{eq:metric_on_sphere}
g_{ij}=\left(\begin{array}{cc} 1 & 0 \\
                       0 & {\rm sin}^2\theta \end{array}\right),\ \ \ \ \ \ \  
g^{ij}=\left(\begin{array}{cc} 1 & 0 \\
                       0 & {\rm sin}^{-2}\theta \end{array}\right) 
\eeq

where the rows and columns are labelled in the order $\theta $, $\phi $.

The distance between points on the symmetric space is defined as
follows.  The length of a vector $X=\sum_it^iX_i$ in the tangent space
${\bf P}$ (this object is well--defined because ${\bf P}$ is endowed
with a definite metric) is identified with the length of the geodesic
connecting the identity element in the coset space with the element
$M={\rm exp}(X)$ \cite{Gilmore}.

\subsection{The algebraic structure of symmetric spaces}
\label{sec-algstr}

Except for the two algebras ${\bf SL(n,R)}$ and ${\bf SU^*(2n)}$ (and
their dual spaces related by the Weyl trick), for which the subspace
representatives of ${\bf K}$, ${\bf P}$ and $i{\bf P}$ consist of
square, irreducible matrices (for ${\bf SL(n,R)}$, we saw this in the
example in subsection \ref{sec-Inv} and for ${\bf SU(n,C)}$ explicitly
in eq.~(\ref{eq:SU3})), the matrix representatives of the subalgebra
${\bf K}$ and of the subspaces ${\bf P}$ and $i{\bf P}$ in the
fundamental representation consist of block--diagonal matrices $X\in
{\bf K}$, $Y\in {\bf P}$, $Y'\in i{\bf P}$ of the form \cite{Gilmore}

\beq
\label{eq:ssrep}
X=\left(\begin{array}{cc} A & 0 \\
                          0 & B \end{array}\right),\ \ \ \ \ \ 
Y=\left(\begin{array}{cc} 0 & C \\
                          -C^\dagger & 0 \end{array}\right),\ \ \ \ \ \
Y'=\left(\begin{array}{cc} 0 & \tilde{C} \\
                          \tilde{C}^\dagger & 0 \end{array}\right), 
\eeq
 
in the Cartan decomposition. Here $A^\dagger =-A$, $B^\dagger =-B$ and
$\tilde{C}=iC$.  In fact, for {\it any} finite--dimensional
representation, the matrix representatives of ${\bf K}$ and ${\bf P}$
are antihermitean (thus they become antisymmetric if the
representation of ${\bf P}$ is real) and as a consequence, those of
$i{\bf P}$ are hermitean (symmetric in case the representation of
$i{\bf P}$ is real) \cite{Gilmore}.  This is true irrespective of
whether the matrix representatives are block--diagonal or square.

The exponential maps of the subspaces ${\bf P}$ and $i{\bf P}$ are
isomorphic to coset spaces $G/K$ and $G^*/K$, respectively (see for
example \cite{Helgason,Hermann}).  The exponential map of the algebra
maps the subspaces ${\bf P}$ and $i{\bf P}$ into unitary and hermitean
matrices, respectively. In the fundamental representation, these
spaces are mapped onto \cite{Gilmore}

\bea
\label{eq:cosetreps}
{\rm exp}({\bf P})={\rm exp}{\left(\begin{array}{cc} 0 & C \\
                                       -C^\dagger & 0 \end{array}\right)}=
\left(\begin{array}{cc} \sqrt{I-XX^\dagger} & X \\
                         -X^\dagger & \sqrt{I-XX^\dagger}\end{array}\right) 
\nonumber \\
{\rm exp}(i{\bf P})={\rm exp}{\left(\begin{array}{cc} 0 & \tilde{C} \\
                                  \tilde{C}^\dagger & 0 \end{array}\right)}=
\left(\begin{array}{cc} \sqrt{I+\tilde{X}\tilde{X}^\dagger} & \tilde{X}\\
    \tilde{X}^\dagger & \sqrt{I+\tilde{X}\tilde{X}^\dagger}\end{array}\right)
\\ \nonumber
\eea

where $X$ is a spherical and $\tilde{X}$ a hyperbolic function of the 
submatrix $C$:

\beq
\label{eq:functions}
X=C\frac{{\rm sin}\sqrt{C^\dagger C}}{\sqrt{C^\dagger C}},\ \ \ 
\tilde{X}=\tilde{C}\frac{{\rm sinh}\sqrt{\tilde{C}^\dagger \tilde{C}}}
{\sqrt{\tilde{C}^\dagger \tilde{C}}}
\eeq

This shows explicitly that the range of parameters parametrizing 
the two cosets is bounded for the compact coset and unbounded for the
non--compact coset, respectively. We already saw an explicit example 
of these formulas in subsection \ref{sec-cosets}.

\section{Real forms of semisimple algebras}
\label{sec-realforms}

In this section we will introduce the tools needed to find all the
real forms of any (semi)simple algebra. The same tools will then be used
in the next section to find the real forms of a symmetric space.
When thinking of a real form, it is convenient to visualize it in
terms of its metric. As we saw in paragraph \ref{sec-metric} the trace
of the metric is called the character of the real form and it
distinguishes the real forms from each other. In the following
subsection we discuss various real forms of an algebra and we see how
to go from one form to another. In each case, we compute the metric
and the character explicitly. We also give the simplest possible example 
of this procedure, the rank--1 algebra. In subsection \ref{sec-realforms2}  
we enumerate the involutive automorphisms needed to classify all real forms
of semisimple algebras and again, we illustrate it with two examples.
 
\subsection{The real forms of a complex algebra}
\label{sec-realforms1}

In general a semisimple complex algebra has several distinct real
forms.  Recall from subsection \ref{sec-rootsp} that a real form of an
algebra is obtained by taking linear combinations of its elements with
real coefficients.  The real forms of the complex Lie algebra ${\bf
  G}$

\beq
\label{eq:complexalg}
\sum_i c^iH_i + \sum_\alpha c^\alpha E_\alpha \ \ \ \ \ \ \ (c^i,\ c^\alpha 
\ {\rm complex}),
\eeq

where ${\bf H_0}=\{ H_i\} $ is the Cartan subalgebra and ${\bf
  G_{\pm \alpha }}= \{E_{\pm \alpha }\}$ are the sets of raising and lowering
operators, can be classified according to all the involutive
automorphisms of ${\bf G}$ obeying $\sigma^2=1$. Two distinctive real
forms are the normal real form and the compact real form.

The {\it normal real form} of the algebra (\ref{eq:complexalg}), which
is also the least compact real form, consists of the subspace in which
the coefficients $c^i$, $c^\alpha$ are real. The metric in this case
with respect to the bases $\{H_i,E_{\pm \alpha}\}$ is (with
appropriate normalization of the elements of the Lie algebra to make
the entries of the metric equal to $\pm 1$)

\beq
\label{eq:raislow_metric}
g_{ij}=\left( \begin{array}{cccccccc}  1 &        &    &   &    &        &   &    \\
                                         & \ddots &    &   &    &        &   &    \\
                                         &        &  1 &   &    &        &   &    \\
                                         &        &    & 0 & 1  &        &   &    \\
                                         &        &    & 1 & 0  &        &   &    \\
                                         &        &    &   &    & \ddots &   &    \\
                                         &        &    &   &    &        & 0 & 1  \\
                                         &        &    &   &    &        & 1 & 0  \end{array} \right)
\eeq

where the $r$ 1's on the diagonal correspond to the elements of the
Cartan subalgebra ($r$ is obviously the rank of the algebra), and the
$2\times 2$ matrices on the diagonal correspond to the pairs $E_{\pm
  \alpha}$ of raising and lowering operators. This structure reflects
the decomposition of the algebra ${\bf G}$ into a direct sum of the
root spaces: ${\bf G}={\bf H_0}\oplus \sum_\alpha {\bf G_\alpha}$.
This metric tensor can be transformed to diagonal form, if we choose
the generators to be

\beq
\label{eq:NRFd}
{\bf K}=\left\{\frac{(E_\alpha - E_{-\alpha})}{\sqrt{2}}\right\},\ \ \ \ \ \ \ 
i{\bf P}=\left\{H_i,\frac{(E_\alpha + E_{-\alpha})}{\sqrt{2}}\right\} 
\eeq

{\bf Example:} In our example with ${\bf SU(3,C)}$, ${\bf K}$ and
$i{\bf P}$ are exactly the subspaces spanned by $\{ X_2,X_5,X_7\} $
and $\{ iX_1,iX_3,iX_4,iX_6,iX_8\}$ (cf. eq.~(\ref{eq:SU3})), and
$(E_\alpha - E_{-\alpha})$ and $-i(E_\alpha + E_{-\alpha})$ are
exactly the Gell--Mann matrices (cf. eq.~(\ref{eq:raislowSU3})).

Then $g_{ij}$ takes the form

\beq
\label{eq:NRFd_metric}
g_{ij}=\left( \begin{array}{cccccccc}  1 &        &    &   &    &        &   &    \\
                                         & \ddots &    &   &    &        &   &    \\
                                         &        &  1 &   &    &        &   &    \\
                                         &        &    & 1 & 0  &        &   &    \\
                                         &        &    & 0 & -1 &        &   &    \\
                                         &        &    &   &    & \ddots &   &    \\
                                         &        &    &   &    &        & 1 & 0  \\
                                         &        &    &   &    &        & 0 & -1  \end{array} \right)
\eeq

where the entries with a minus sign correspond to the generators of
the compact subalgebra ${\bf K}$, the first $r$ entries equal to $+1$ 
correspond to the Cartan subalgebra, and the remaining ones to the operators
in $i{\bf P}$ {\it not} in the Cartan subalgebra.  
This is the diagonal metric tensor corresponding to 
the normal real form. The character of the normal real form is plus the rank 
of the algebra. 

The {\it compact real form} of ${\bf G}$ is obtained 
from the normal real form by the Weyl unitary trick:

\beq
\label{eq:compactrealform}
{\bf K}=\left\{\frac{(E_\alpha - E_{-\alpha})}{\sqrt{2}}\right\},\ \ \ \ \ \ \ 
{\bf P}=\left\{iH_i,\frac{i(E_\alpha + E_{-\alpha})}{\sqrt{2}}\right\} 
\eeq

The character of the compact real form is minus the dimension of the
algebra, and the metric tensor is $g_{ij}={\rm diag}(-1,...,-1)$.  

{\bf Example:} We will use as an example the well--known ${\bf
  SU(2,C)}$ algebra with Cartan subalgebra ${\bf H_0}=\{ J_3\}$ and
raising and lowering operators $\{ J_\pm \} $.  We have chosen the
normalization such that the non--zero entries of $g_{ij}$ are all
equal to $1$:

\beq
\begin{array}{c}
J_3=\frac{1}{2\sqrt{2}}\tau_3,\ \ \ \ \ \ \ 
J_\pm =\frac{1}{4}(\tau_1\pm i\tau_2)\end{array}
\eeq

where in the defining representation of ${\bf SU(2,C)}$

\beq
\tau_3= \left( \begin{array}{cc} 1 & 0  \\ 
                                   0 & -1 \end{array} \right),
\ \ \ \ 
\tau_1= \left( \begin{array}{cc} 0 & 1  \\ 
                                   1 & 0  \end{array} \right),
\ \ \ \ 
\tau_2= \left( \begin{array}{cc} 0 & -i \\ 
                                   i & 0  \end{array} \right)
\eeq

The normalization is such that 

\beq
\begin{array}{c}
[J_3,J_\pm]=\pm \frac{1}{\sqrt{2}}J_\pm,\ \ \ \ \ \ \ 
[J_+,J_-]=\frac{1}{\sqrt{2}}J_3\end{array}
\eeq

In equation (\ref{eq:SU2adjoint}) we constructed the adjoint
representation of this algebra, albeit with a different normalization.
Using the present normalization to set the entries of the metric equal
to 1, we see that the non--zero structure constants are
$C^+_{3+}=-C^+_{+3}=-C^-_{3-}=C^-_{-3}=C^3_{+-}=-C^3_{-+}=\frac{1}{\sqrt{2}}$.
The entries of the metric are given by eq.~(\ref{eq:metric}),
$g_{ij}=K(J_i,J_j)=C^r_{is}C^s_{jr}$ with summation over repeated
indices, so we see that the metric of the normal real form ${\bf
  SU(2,R)}$ in this basis is

\beq
g_{ij}=\left( \begin{array}{ccc} 1 & 0 & 0 \\
                                 0 & 0 & 1 \\
                                 0 & 1 & 0 \end{array} \right)
\eeq

where the rows and columns are labelled by $3,+,-$ respectively. This 
corresponds to eq.~(\ref{eq:raislow_metric}).

To pass now to a diagonal metric, we just have to set

\beq
\begin{array}{c}
\Sigma_3=J_3 \nonumber \\ \nonumber \\
\Sigma_1=\frac{J_++J_-}{\sqrt{2}}=\frac{1}{2\sqrt 2}\tau_1 \nonumber \\
\nonumber \\
\Sigma_2=\frac{J_+-J_-}{\sqrt{2}}=\frac{i}{2\sqrt 2}\tau_2 \end{array} 
\nonumber 
\eeq

like in eq.~(\ref{eq:NRFd}). The commutation relations then become

\beq
\label{eq:Sigmarels}
\begin{array}{c} 
[\Sigma_1,\Sigma_2]=-\frac{1}{\sqrt 2}\Sigma_3,\ \ \ \ \ \ \ 
[\Sigma_2,\Sigma_3]=-\frac{1}{\sqrt 2}\Sigma_1,\ \ \ \ \ \ \ 
[\Sigma_3,\Sigma_1]=\frac{1}{\sqrt 2}\Sigma_2\end{array}
\eeq

These commutation relations characterize the algebra ${\bf
  SO(2,1;R)}$.  From here we find the structure constants
$C^3_{12}=-C^3_{21}=C^1_{23}=-C^1_{32}=-C^2_{31}=C^2_{13}=
-\frac{1}{\sqrt 2}$ and the diagonal metric of the normal real form
with rows and columns labelled $3,1,2$ (in order to comply with the
notation in eq.~(\ref{eq:NRFd_metric})) is

\beq
\label{eq:SO21}
g_{ij}=\left( \begin{array}{ccc} 1 & 0 & 0 \\
                                 0 & 1 & 0 \\
                                 0 & 0 & -1 \end{array} \right)
\eeq

which is to be compared with eq.~(\ref{eq:NRFd_metric}).  According to
eq.~(\ref{eq:NRFd}), the Cartan decomposition of ${\bf G^*}$ is ${\bf
  G^*}={\bf K}\oplus i{\bf P}$ where ${\bf K}=\{\Sigma_2\}$ and $i{\bf
  P}=\{\Sigma_3,\Sigma_1\}$.  The Cartan subalgebra of $i{\bf P}$
consists of $\Sigma_3$.

Finally, we arrive at the compact real form by multiplying $\Sigma_3$
and $\Sigma_1$ with $i$.  Setting $i\Sigma_1=\tilde{\Sigma}_1$,
$\Sigma_2=\tilde{\Sigma}_2$, $i\Sigma_3=\tilde{\Sigma}_3$ the
commutation relations become those of the special orthogonal group:

\beq
\label{eq:tildeSigmarels}
\begin{array}{c} 
[\tilde{\Sigma}_1,\tilde{\Sigma}_2]=-\frac{1}{\sqrt 2}\tilde{\Sigma}_3,\ \ \ \ \ \ \ 
[\tilde{\Sigma}_2,\tilde{\Sigma}_3]=-\frac{1}{\sqrt 2}\tilde{\Sigma}_1,\ \ \ \ \ \ \ 
[\tilde{\Sigma}_3,\tilde{\Sigma}_1]=-\frac{1}{\sqrt 2}\tilde{\Sigma}_2\end{array}
\eeq

The last commutation relation in eq.~(\ref{eq:Sigmarels}) has changed
sign whereas the others are unchanged.  $C^2_{31}$, $C^2_{13}$, and
consequently $g_{33}$ and $g_{11}$ change sign and we get the metric 
for ${\bf SO(3,R)}$:

\beq
\label{eq:SO3}
g_{ij}=\left( \begin{array}{ccc} -1 & 0 & 0 \\
                                 0 & -1 & 0 \\
                                 0 & 0 & -1 \end{array} \right)
\eeq 

This is the compact real form. The subspaces of the compact algebra
${\bf G}={\bf K}\oplus {\bf P}$ are ${\bf K}=\{\tilde{\Sigma}_2\}$ and
${\bf P}=\{\tilde{\Sigma}_3,\tilde{\Sigma}_1\}$.  Weyl's theorem
states that a simple Lie group $G$ is compact, if and only if the
Killing form on ${\bf G}$ is negative definite; otherwise it is
non--compact.  In the present example, we see this explicitly.

\subsection{The classification machinery}
\label{sec-realforms2}

To classify all the real forms of any complex Lie algebra, with
characters lying between the character of the normal real form and
the compact real form (the intermediate real forms obviously have an
indefinite metric), it suffices to enumerate all the involutive
automorphisms of its compact real form.  A detailed and almost
complete account of these procedures for the non--exceptional groups
can be found in \cite{Gilmore}, Chapter~9, paragraph~3. To summarize,
if ${\bf G}$ is the compact real form of a complex semisimple Lie
algebra ${\bf G^C}$, ${\bf G^*}$ runs through all its associated 
non--compact real
forms ${\bf G^*}$, ${\bf G'^*}$, ... with corresponding maximal
compact subgroups ${\bf K}$, ${\bf K'}$, ...  and complementary
subspaces $i{\bf P}$, $i{\bf P'}$, ...  as $\sigma $ runs through all
the involutive automorphisms of ${\bf G}$.

One such automorphism is complex conjugation $\sigma_1 =K$, which is
used to split the compact real algebra into subspaces ${\bf K}$ and
${\bf P}$ in eq.~(\ref{eq:compactrealform}). (To avoid confusion: the
generators can be complex even though the field of real numbers is
used to multiply the generators in a real form of an algebra. If the
generators are also real, we speak of a real representation. However,
whether we consider the field to be ${\bf R}$ and the generators to be
complex, or the opposite, also depends on our definition of basis.)
The involutive automorphisms $\sigma $ satisfy $\sigma {\bf G}
\sigma^{-1} = {\bf G} $, $\sigma^2=1$, which implies that $\sigma $
either commutes or anticommutes with the elements of the compact
algebra ${\bf G}$: if $\sigma X \sigma^{-1} = X'$, then $\sigma X'
\sigma^{-1} = X$, and we get $X'=\pm X$ for $X,\ X'\in {\bf G}$ (see
the example below).  One can show \cite{Loos} (Ch.~VII), that it
suffices to consider the following three possibilities for $\sigma $:
$\sigma_1 =K$, $\sigma_2 = I_{p,q}$ and $\sigma_3 = J_{p,p}$ where

\beq
I_{p,q} = \left( \begin{array}{cc} I_p & 0 \\
                                   0  & -I_q \end{array} \right), 
\ \ \ \ \ \  J_{p,p} = \left( \begin{array}{cc} 0 & I_p \\
                                   -I_p  &  0 \end{array} \right) 
\eeq

and $I_p$ denotes the $p \times p$ unit matrix. By operating with one
(or two successive ones) of these automorphisms on the elements of
${\bf G}$, we can construct the subspaces ${\bf K}$ and ${\bf P}$, and
${\bf K}$ and $i{\bf P}$ of the corresponding non--compact real form
${\bf G^*}$.  A complex algebra and all its real forms (the compact
and the various non--compact ones) correspond to the same root lattice
and Dynkin diagram.

{\bf Example}: The normal real form of the complex algebra ${\bf
  G^C}={\bf SL(n,C)}$ is the non--compact algebra ${\bf G^*}={\bf
  SL(n,R)}$. As we saw in subsection \ref{sec-Inv}, this algebra can
be decomposed as ${\bf K}\oplus i{\bf P}$ where ${\bf K}$ is the
algebra consisting of real, skew--symmetric and traceless $n\times n$
matrices and $i{\bf P}$ is the algebra consisting of real, symmetric
and traceless $n\times n$ matrices.  Under the Weyl unitary trick we
constructed, in a previous example, this algebra from the compact real
form of ${\bf G^C}$, ${\bf SU(n,C)}={\bf G}={\bf K}\oplus {\bf P}$.

Starting with the compact real form ${\bf G}$, we can construct all
the various non--compact real forms ${\bf G^*}$, ${\bf G'^*}$,...
from it, by applying the involutive automorphisms $\sigma_1$,
$\sigma_2$, $\sigma_3$ to the elements of ${\bf G}$.  All the real
forms related to the root system $A_{n-1}$ are obtained by applying
the three involutions to ${\bf G}={\bf SU(n,C)}$:
 
$\sigma_1$) The involutive automorphism $\sigma_1=K$ (complex
conjugation) splits ${\bf G}={\bf SU(n,C)}$ into ${\bf K}\oplus {\bf P}$ 
(we recall this from the example in paragraph \ref{sec-Inv}).
The non--compact real form obtained this way, by the Weyl unitary
trick, is exactly the normal real form ${\bf G^*}={\bf K}\oplus 
i{\bf P}= {\bf SL(n,R)}$.

$\sigma_2$) A general matrix in the Lie algebra ${\bf SU(n,C)}$ can be
written in the form 

\beq
\label{eq:generalSU(n)}
X=\left( \begin{array}{cc} A & B \\
                          -B^\dagger & C \end{array}\right)
\eeq

where $A$, $C$ are complex $p\times p$ and $q\times q$ matrices
satisfying $A^\dagger = -A$, $C^\dagger = -C$, ${\rm tr}A+{\rm tr}C=0$
(since the determinant of the group elements must be $+1$), and $B$ is
an arbitrary complex $p\times q$ matrix ($p+q=n$). In
eq.~(\ref{eq:generalSU(n)}), the matrices $A$, $B$ and $C$ are all
linear combinations of submatrices in {\it both} subspaces ${\bf
  K}=\{\frac{1}{\sqrt{2}}(E_{\alpha_i}-E_{-\alpha_i})\}$ and ${\bf
  P}=\{iH_j,\frac{i}{\sqrt{2}}(E_{\alpha_i}+E_{-\alpha_i})\}$. The
action of the involution $\sigma_2=I_{p,q}$ on $X$ is

\beq
I_{p,q}XI_{p,q}^{-1}= \left( \begin{array}{cc} A & -B \\
                          B^\dagger & C \end{array}\right)
\eeq

Therefore, we see that the subspaces ${\bf K'}$ and ${\bf P'}$ are given by
the matrices

\beq
\left( \begin{array}{cc} A & 0 \\
                          0 & C \end{array}\right) \in {\bf K'},
\ \ \ \ \ \ \ 
\left( \begin{array}{cc} 0 & B \\
                          -B^\dagger & 0 \end{array}\right) \in {\bf P'}
\eeq

Indeed, we see that $I_{p,q}$ transforms the Lie algebra elements in 
${\bf K'}$ into themselves, and those in ${\bf P'}$ into minus themselves.
The transformation by $I_{p,q}$ mixes the subspaces ${\bf K}$ and ${\bf P}$,
and splits the algebra in a different way into ${\bf K'}\oplus {\bf P'}$. 
The matrices

\beq
\left( \begin{array}{cc} A & iB \\
                         -iB^\dagger & C \end{array}\right)
\in {\bf K'}\oplus i{\bf P'}
\eeq

define the non--compact real form ${\bf G'^*}$. 
This algebra is called ${\bf SU(p,q;C)}$ and its maximal 
compact subalgebra ${\bf K'}$ is ${\bf SU(p)\otimes SU(q)\otimes U(1)}$. 

$\sigma_3$) By the involutive automorphism $\sigma_3\sigma_1=J_{p,p}K$
one constructs in a similar way (for details see \cite{Gilmore}) a
third non--compact real form (for even $n=2p$) ${\bf G''^*}={\bf
  K''}\oplus i{\bf P''}$ associated to the algebra ${\bf G}={\bf
  SU(2p,C)}$.  ${\bf G''^*}$ is the algebra ${\bf SU^*(2p)}$ and its
maximal compact subalgebra is ${\bf USp(2p)}$.  \footnote{ The algebra
  ${\bf SU^*(2p)}$ is represented by complex $2p\times 2p$ matrices of
  the form \beq
  X=\left(\begin{array}{cc} A & B \\
      -B^* & -A^*\end{array}\right) \eeq where ${\rm tr}A+{\rm
    tr}A^*=0$.  ${\bf USp(2p)}$ denotes the complex $2p\times 2p$
  matrix algebra of the group with both unitary and symplectic
  symmetry (${\bf USp(2p,C)}$ can also be denoted ${\bf U(p,Q)}$ where
  $Q$ is the field of quaternions).  A matrix in the algebra ${\bf
    USp(2p,C)}$ can be written as \beq
  X=\left(\begin{array}{cc} A & B \\
      -B^{\dagger } & -A^R\end{array}\right) \eeq where
  $A^\dagger=-A$, $B^R=B$, and the superscript $^R$ denotes reflection
  in the minor diagonal. }

This procedure, summarized in the formula below, 
exhausts all the real forms of the simple algebras.  

\beq
{\bf G^C} \to {\bf G}={\bf K}\oplus {\bf P}  
\begin{array}{l}{\buildrel{\scriptstyle \sigma_1} \over \nearrow }\\
 {\buildrel{\scriptstyle \sigma_2} \over \to } \\
 {\buildrel{\scriptstyle \sigma_3} \over \searrow }
\end{array}
\begin{array}{l}
{\bf G^*}={\bf K}\oplus i{\bf P}\\ \\
{\bf G'^*}={\bf K'}\oplus i{\bf P'}\\ \\
{\bf G''^*}={\bf K''}\oplus i{\bf P''}
\end{array}
\eeq

{\bf Example:} Note that it may not always be possible to apply all
the above involutions $\sigma_1$, $\sigma_2$, $\sigma_3$ to the
algebra. For example, complex conjugation $\sigma_1$ does not do
anything to ${\bf SO(2n+1,R)}$, because it is represented by real
matrices, neither is $\sigma_3$ a symmetry of this algebra, since the
adjoint representation is odd--dimensional and $\sigma_3$ has to act
on a $2p\times 2p$ matrix. The only possibility remains
$\sigma_2=I_{p,q}$.  For a second, even more concrete example, let's
look at the algebra ${\bf SO(3,R)}$, belonging to the root lattice
$B_1$. This algebra is spanned by the generators $L_1$, $L_2$, $L_3$
given in subsection \ref{sec-cosets}. A general element of the algebra
is

\beq
X={\bf t\cdot L}=\frac{1}{2}\left(\begin{array}{ccc}     & t^3 & t^2\\
                   -t^3 &     & t^1\\
                   -t^2 & -t^1& \end{array}\right)=
\frac{1}{2}\left(\begin{array}{ccc}     & t^3 & \\
                   -t^3 &     & \\
                        &     & \end{array}\right) \oplus
\frac{1}{2}\left(\begin{array}{ccc}     &     & t^2\\
                        &     & t^1\\
                   -t^2 & -t^1&    \end{array}\right)
\eeq

This splitting of the algebra is caused by the involution $I_{2,1}$ acting
on the representation:

\beq
I_{2,1}XI_{2,1}^{-1}=\left(\begin{array}{ccc} 1 &    &    \\
                                           &  1 &    \\
                                           &    & -1 \end{array}\right)
\frac{1}{2}\left(\begin{array}{ccc}     & t^3 & t^2\\
                                   -t^3 &     & t^1\\
                                   -t^2 & -t^1& \end{array}\right)
\left(\begin{array}{ccc} 1 &    &    \\
                           &  1 &    \\
                           &    & -1 \end{array}\right)
=\frac{1}{2}\left(\begin{array}{ccc}     & t^3 & -t^2\\
                                    -t^3 &     & -t^1\\
                                     t^2 & t^1& \end{array}\right)
\eeq

and it splits it into 
${\bf SO(3)}={\bf K}\oplus {\bf P}={\bf SO(2)}\oplus {\bf SO(3)/SO(2)}$.
Exponentiating, as we saw in subsection \ref{sec-cosets}, the coset 
representative is a point on the 2--sphere

\beq
M=\left(\begin{array}{ccc}
.&.&t^2\frac{{\rm sin}\sqrt{(t^1)^2+(t^2)^2}}{\sqrt{(t^1)^2+(t^2)^2}}\\
.&.&t^1\frac{{\rm sin}\sqrt{(t^1)^2+(t^2)^2}}{\sqrt{(t^1)^2+(t^2)^2}}\\
.&.&{\rm cos}\sqrt{(t^1)^2+(t^2)^2}\end{array}\right)=
\left(\begin{array}{ccc} . & .  & x\\
                         . & .  & y\\
                         . &  . & z   \end{array}\right);\ \ \ \ \ \ \  
x^2+y^2+z^2=1
\eeq

By the Weyl unitary trick we now get the non--compact real form 
${\bf G^*}={\bf K}\oplus i{\bf P}$: 
${\bf SO(2,1)}={\bf SO(2)}\oplus {\bf SO(2,1)/SO(2)}$. This algebra is
represented by 

\beq
\label{eq:SO(2,1)}
\left(\begin{array}{ccc}     & t^3 & it^2\\
                   -t^3 &     & it^1\\
                   -it^2 & -it^1& \end{array}\right)=
\left(\begin{array}{ccc}     & t^3 & \\
                   -t^3 &     & \\
                        &     & \end{array}\right) \oplus
\left(\begin{array}{ccc}     &     & it^2\\
                        &     & it^1\\
                   -it^2 & -it^1&    \end{array}\right)
\eeq

and after exponentiation of the coset generators

\beq
M=\left(\begin{array}{ccc} . & . & it^2\frac{{\rm sinh}\sqrt{(t^1)^2+(t^2)^2}}
                             {\sqrt{(t^1)^2+(t^2)^2}} \\
                      . & . & it^1\frac{{\rm sinh}\sqrt{(t^1)^2+(t^2)^2}}
                             {\sqrt{(t^1)^2+(t^2)^2}}\\
                      . & . & {\rm cosh}\sqrt{(t^1)^2+(t^2)^2}  \end{array}\right)=
\left(\begin{array}{ccc} .  & .  & ix\\
                         .  & .  & iy\\
                         .  & .  & z   \end{array}\right);\ \ \ \ \ \ \  
(ix)^2+(iy)^2+z^2=1
\eeq

The surface in ${\bf R}^3$ consisting of points $(x,y,z)$ satisfying this
equation is the hyperboloid $H^2$. Similarly, we get the isomorphic space
$SO(1,2)/SO(2)$ by applying $I_{1,2}$: ${\bf SO(1,2)}={\bf \tilde{K}}\oplus 
i{\bf \tilde{P}}={\bf SO(2)}\oplus {\bf SO(1,2)/SO(2)}$ and in terms of the 
algebra

\beq
\tilde{X}=
\frac{1}{2}\left(\begin{array}{ccc}  &      &     \\
                                     &      & t^1 \\
                                     & -t^1 &     \end{array}\right) \oplus
\frac{1}{2}\left(\begin{array}{ccc}     & -it^3 & -it^2\\
                                  it^3  &       &      \\
                                  it^2  &       & \end{array}\right)
\eeq

\section{The classification of symmetric spaces}
\label{sec-claSS}

In this section we introduce the curvature tensor and the sectional
curvature of a symmetric space, and we extend the family of symmetric
spaces to include also flat or Euclidean--type spaces. These are
identified with the subspace ${\bf P}$ of the Lie algebra itself, and
the group that acts on it is a semidirect product of the subgroup $K$
and the subspace ${\bf P}$. As we will learn, to each compact subgroup
$K$ corresponds a triplet of symmetric spaces with positive, zero and
negative curvature.  The classification of these symmetric spaces is
in exact correspondence with the new classification of random matrix
models to be discussed in \cite{CasMag}. These spaces exhaust the
Cartan classification and have a definite metric. They are listed in
Table~1 together with some of their properties.

In paragraph \ref{sec-restricted} we introduce restricted root
systems.  In the same way as a Lie algebra corresponds to a given root
system, each symmetric space corresponds to a restricted root system.
These root systems are of primary importance in the physical
applications to be discussed in our forthcoming paper \cite{CasMag}.
The restricted root system can be of an entirely different type from
the root system inherited from the complex extension algebra, and its
rank may be different. We work out a specific example of a restricted
root system as an illustration. In spite of their importance, we have
not been able to find any explicit reference in the literature that
explains how to obtain the restricted root systems. Instead, we found
that they are often referred to in tables and in mathematical texts
without explicitly mentioning that they are restricted, which could
easily lead to confusion with the inherited root systems. In reference
\cite{Gilmore} the root system that is associated to each symmetric
space is the one inherited from the complex extension algebra, whereas
for example in Table~B1 of reference~\cite{OlshPere} and in
\cite{Loos} the restricted root systems are listed.
 
There are also symmetric spaces with an indefinite metric, so called
pseudo--Riemannian spaces, corresponding to a maximal {\it
  non--compact} subgroup $H$. For completeness, we will briefly
discuss how these are obtained as real forms of symmetric spaces
corresponding to compact symmetric subgroups.  This does not require
any new tools than the ones we have already introduced, namely the
involutive automorphisms.

\subsection{The curvature tensor and triplicity}
\label{sec-curv}

Suppose that ${\bf K}$ is a maximal compact subalgebra of the
non--compact algebra ${\bf G^*}$ in the Cartan decomposition ${\bf
  G^*} = {\bf K} \oplus i{\bf P}$, where $i{\bf P}$ is a complementary
subspace. ${\bf K}$ and ${\bf P}$ (alternatively ${\bf K}$ and $i{\bf
  P}$) satisfy eq.~(\ref{eq:commrel}):

\beq
\label{eq:commrel2}
[{\bf K},{\bf K}]\subset {\bf K},\ \  [{\bf K},{\bf P}]\subset {\bf P},\ \  
[{\bf P},{\bf P}]\subset {\bf K} 
\eeq

${\bf K}$ is called a symmetric subalgebra and the coset spaces $\exp
({\bf P})\simeq G/K$ and $\exp (i{\bf P})\simeq G^*/K$ are globally
symmetric Riemannian spaces. Globally symmetric means that every point
on the manifold can be moved to any other point by a particular group
operation (we discussed this in paragraph \ref{sec-cosets}; for a
rigorous definition of globally symmetric spaces see Helgason
\cite{Helgason}, paragraph IV.3).  In the same way, the metric can be
defined in any point of the manifold by moving the metric at the
origin to this point, using a group operation (cf. eq.~(\ref{eq:g(M)})
in paragraph \ref{sec-metric}).  The Killing form restricted to the
tangent spaces ${\bf P}$ and $i{\bf P}$ at any point in the coset
manifold has a definite sign. The manifold is then called
``Riemannian''. The metric can be taken to be either plus or minus the
Killing form so that it is always positive definite (cf.
paragraph~\ref{sec-metric}).
 
A {\it curvature tensor} with components $R^i_{jkl}$ can be defined on
the manifold $G/K$ or $G^*/K$ in the usual way
\cite{Helgason,FosNigh}. It is a function of the metric tensor and its
derivatives. It was proved for instance in \cite{Helgason}, Ch. IV,
that the components of the curvature tensor at the origin of a
globally symmetric coset manifold is given by the expression

\beq
\label{eq:curv}
R^n_{ijk}X_n= [X_i,[X_j,X_k]] = C^n_{im}C^m_{jk}X_n
\eeq

where $\{X_i\}$ is a basis for the Lie algebra. The sectional
curvature at a point $p$ is equal to

\beq
\label{eq:K}
{\cal K}=g([[X,Y],X],Y)
\eeq

where $g$ is an arbitrary symmetric and nondegenerate metric (such a
metric is also called a pseudo--Riemannian structure, or simply a
Riemannian structure if it has a definite sign) on the tangent
space at $p$, invariant under the action of the group elements.  In
(\ref{eq:K}), $g(X_i,X_j)\equiv g_{ij}$ and $\{ X,Y\} $ is an
orthonormal basis for a two--dimensional subspace $S$ of the tangent
space at the point $p$ (assuming it has dimension $\geq 2$).  The
sectional curvature is equal to the gaussian curvature on a
2--dimensional manifold. If the manifold has dimension $\geq 2$,
(\ref{eq:K}) gives the sectional curvature along the section $S$.

Eqs.~(\ref{eq:curv}) and (\ref{eq:K}), together with
eq.~(\ref{eq:commrel2}) show that the curvature of the spaces $G/K$
and $G^*/K$ has a definite and opposite sign (\cite{Helgason},
par.~V.3). Thus, we see that if $G$ is a compact semisimple group, to
the same subgroup $K$ there corresponds a positive curvature space
$P\simeq G/K$ and a dual negative curvature space $P^*\simeq G^*/K$.
The reason for this is exactly the same as the reason why the sign changes 
for the components of the metric corresponding to the generators in $i{\bf P}$
as we go to the dual space ${\bf P}$. We remind the reader that the sign of the
metric can be chosen positive or negative for a compact space. The issue here
is that the sign changes in going from $G^*/K$ to $G/K$. 

{\bf Example:} We can use the example of $SU(2)$ in paragraph
\ref{sec-realforms1} to see that the sectional curvature is the
opposite for the two spaces $G/K$ and $G^*/K$. If we take $\{
X,Y\}=\{\Sigma_3, \Sigma_1\}$ as the basis in the space $i{\bf P}$ and
$\{\tilde{\Sigma}_3, \tilde{\Sigma}_1\}$ ($\tilde{\Sigma}_i\equiv
i\Sigma_i$) as the basis in the space ${\bf P}$, we see by comparing
the signs of the entries of the metrics we computed in eqs.~(\ref{eq:SO21})
and (\ref{eq:SO3}) that the sectional curvature ${\cal K}$ at the origin 
has the opposite sign for the two spaces $SO(2,1)/SO(2)$ and $SO(3)/SO(2)$.

Actually, there is also a zero--curvature symmetric space $X^0=G^0/K$
related to $X^+=G/K$ and $X^-=G^*/K$, so that we can speak of a {\it
  triplet} of symmetric spaces related to the {\it same} symmetric
subgroup $K$. The zero--curvature spaces were discussed in
\cite{OlshPere} and in Ch.~V of Helgason's book \cite{Helgason}, where
they are referred to as ``symmetric spaces of the euclidean type''.
That their curvature is zero was proved in Theorem~3.1 of
\cite{Helgason}, Ch.~V.

The flat symmetric space $X^0$ can be identified with
the subspace ${\bf P}$ of the algebra. The group $G^0$ is a semidirect
product of the subgroup $K$ and the invariant subspace ${\bf P}$ of
the algebra, and its elements $g=(k,a)$ act on the elements of $X^0$
in the following way:

\beq
g(x)=kx+a,\ \ \ \ k\in K,\ \ \ \  x,a \in X^0
\eeq

if the $x$'s are vectors, and 

\beq
g(x)=kxk^{-1}+a,\ \ \ \ k\in K,\ \ \ \  x,a \in X^0
\eeq

if the $x$'s are matrices. We will see one example of each below.

The elements of the algebra ${\bf P}$ now define an {\it abelian
  additive group}, and $X^0$ is a vector space with euclidean
geometry.  In the above scenario, the subspace ${\bf P}$ contains only
the operators of the Cartan subalgebra and no others: ${\bf P}={\bf
  H_0}$, so that ${\bf P}$ is a subalgebra of ${\bf G^0}$. The algebra
${\bf G^0}={\bf K}\oplus {\bf P}$ belongs to a non--semisimple group
$G^0$, since it has an abelian ideal ${\bf P}$: $[{\bf K},{\bf
  K}]\subset {\bf K}$, $[{\bf K},{\bf P}]\subset {\bf P}$, $[{\bf
  P},{\bf P}]=0$. Note that ${\bf K}$ and ${\bf P}$ still satisfy the
commutation relations (\ref{eq:commrel2}).  In this case the coset
space $X^0$ is flat, since by (\ref{eq:commrel2}), $R^n_{ijk}=0$ for
all the elements $X\in {\bf P}$.  Eq.~(\ref{eq:curv}) is valid for any
space with a Riemannian structure.  Indeed, it is easy to see from
eqs. (\ref{eq:curv}), (\ref{eq:K}) that $R^n_{ijk}={\cal K}=0$ if the
generators are abelian.  Even though the non--semisimple algebras have
a degenerate metric, it is trivial to find a non--degenerate metric on
the symmetric space $X^0$ that can be used in ~(\ref{eq:K}) to find
that the sectional curvature at any point is zero.  For example, as we
pass from the sphere to the plane, the metric becomes degenerate in
the limit as $[L_1,L_2]\sim L_3\to [P_1,P_2]=0$ (see the example
below). Obviously, we do not inherit this degenerate metric from the
tangent space on ${\bf R^2}$ like in the case of the sphere, but the
usual metric for ${\bf R^2}$, $g_{ij}=\delta_{ij}$ provides the
Riemannian structure on the plane.

{\bf Examples:} An example of a flat symmetric space is $E_2/K$, where
$G^0=E_2$ is the euclidean group of motions of the plane ${\bf R^2}$:
$g(x)= kx+a$, $g=(k,a)\in G^0$ where $k\in K=SO(2)$ and $a\in {\bf
  R}^2$.  The generators of this group are translations $P_1$, $P_2
\in {\bf H_0}= {\bf P}$ and a rotation $J\in {\bf K}$ satisfying
$[P_1,P_2]=0$, $[J,P_i]=-\epsilon^{ij}P_j$, $[J,J]=0$, in agreement
with eq.~(\ref{eq:commrel2}) defining a symmetric subgroup. The
abelian algebra of translations $\sum_{i=1}^2t^iP_i$, $t^i\in {\bf
  R}$, is isomorphic to the plane ${\bf R^2}$, and can be identified
with it.

The commutation relations for $E_2$ are a kind of limiting case of the
commutation relations for ${\bf SO(3)} \sim {\bf SU(2)}$ and ${\bf
  SO(2,1)}$. If in the limit of infinite radius of the sphere $S^2$ we
identify $\tilde{\Sigma}_1$ with $P_1$, $\tilde{\Sigma}_2$ with $P_2$,
and $\tilde{\Sigma}_3$ with $J$, we see that the commutation relations
resemble the ones described in eq.~(\ref{eq:Sigmarels}) and
(\ref{eq:tildeSigmarels}) -- we only have to set
$[\tilde{\Sigma}_1,\tilde{\Sigma}_2]=0$, which amounts to setting
$C_{12}^3=-C_{21}^3\to 0$. From here we get the degenerate metric of
the non--semisimple algebra ${\bf E_2}$:

\beq
g_{ij}=\left(\begin{array}{ccc} -1 & & \\ & 0 & \\ & & 0 \end{array}\right)
\eeq

where the only nonzero element is $g_{33}$. This is to be confronted
with eqs.~(\ref{eq:SO21}) and (\ref{eq:SO3}) which are the metrics for
${\bf SO(2,1)}$ and ${\bf SO(3)}$. This is an example of contraction of 
an algebra.

An example of a triplet $\{X^+, X^0, X^-\}$ corresponding to the same 
subgroup $K=SO(n)$ is: 

1) $X^+=SU(n,C)/SO(n)$, the set of symmetric unitary matrices with
unit determinant; it is the space ${\rm exp}({\bf P})$ where ${\bf P}$
are real, symmetric and traceless $n \times n$ matrices.  (Cf. the
example in subsection \ref{sec-Inv}.)

2) $X^0$ is the set ${\bf P}$ of real, symmetric and traceless
$n\times n$ matrices and the non--semisimple group $G^0$ is the group
whose action is defined by $g(x)=kxk^{-1}+a$, $g=(k,a)\in G^0$ where
$k\in K=SO(n)$ and $x,a\in X^0$.  The involutive automorphism maps
$g=(k,a)\in G^0$ into $g'=(k,-a)$.

3) $X^-=SL(n,R)/SO(n)$ is the set of real, positive, symmetric
matrices with unit determinant; it is the space ${\rm exp}(i{\bf P})$
where ${\bf P}$ are real, symmetric and traceless $n \times n$
matrices.

We remark that the zero--curvature symmetric spaces correspond to the
integration manifolds of many known matrix models with physical
applications.

The pairs of dual symmetric spaces of positive and negative curvature
listed in each row of Table~1 originate in the same complex extension
algebra \cite{Gilmore} with a given root lattice. This ``inherited''
root lattice is listed in the first column of the table.  In our
example in paragraph \ref{sec-realforms2} this was the root lattice of
the complex algebra ${\bf G^C}= {\bf SL(n,C)}$. The same root lattice
$A_{n-1}$ characterizes the real forms of ${\bf SL(n,C)}$: as we saw
in the example these are the algebras ${\bf SU(n,C)}$, ${\bf
  SL(n,R)}$, ${\bf SU(p,q;C)}$ and ${\bf SU^*(2n)}$, and we have seen
how to construct them using involutive automorphisms.

However, also listed in Table~1 is the {\it restricted root system}
corresponding to each symmetric space. This root system may be
different from the one inherited from the complex extension algebra.
Below, we will define the restricted root system and see an explicit
example of one such system. While the original root lattice
characterizes the complex extension algebra and its real forms, the
restricted root lattice characterizes a particular symmetric space
originating from one of its real forms.  The root lattices of the
classical simple algebras are the infinite sequences $A_n$, $B_n$,
$C_n$, $D_n$, where the index $n$ denotes the rank of the
corresponding group.  The {\it root multiplicities} $m_o$, $m_l$,
$m_s$ listed in Table~1 (where the subscripts refer to ordinary, long
and short roots, respectively) are characteristic of the restricted
root lattices.  In general, in the root lattice of a simple algebra
(or in the graphical representation of any irreducible
representation), the roots (weights) may be degenerate and thus have a
multiplicity greater than 1. This happens if the same weight $\mu =
(\mu_1,...,\mu_r)$ corresponds to different states in the
representation. In that case one can arrive at that particular weight
using different sets of lowering operators $E_{-\alpha}$ on the
highest weight of the representation. Indeed, we saw in the example of
$SU(3,C)$ in subsection \ref{sec-rootsp}, that the roots can have a
multiplicity different from 1. The same is true for the restricted
roots.

The sets of simple roots of the classical root systems (briefly listed
in subsection \ref{sec-rootsp}) have been obtained for example in
\cite{Gilmore,Georgi}. In the canonical basis in ${\bf R^n}$, the
roots of type $\{ \pm e_i \pm e_j, i\neq j\}$ are ordinary while the
roots $\{ \pm 2e_i \}$ are long and the roots $\{ \pm e_i \}$ are
short.  Only a few sets of root multiplicities are compatible with the
strict properties characterizing root lattices in general.

\subsection{Restricted root systems}
\label{sec-restricted}

The restricted root systems play an important role in connection with
matrix models and integrable Calogero--Sutherland models. We will
discuss this in detail in \cite{CasMag}. In this subsection we will
explain how restricted root systems are obtained and how they are
related to a given symmetric space.\footnote{The author is indebted to
  Prof. Simon Salamon for explaining to her how the restricted root
  systems are obtained.}

As we have repeatedly seen in the examples using the compact algebra
${\bf SU(n,C)}$ (in particular in subsection \ref{sec-realforms2}),
the algebra ${\bf SU(p,q;C)}$ ($p+q=n$) is a non--compact real form of
the former. This means they share the same rank--$(n-1)$ root system
$A_{n-1}$. However, to the {\it symmetric space}
$SU(p,q;C)/(SU(p)\otimes SU(q)\otimes U(1))$ one can associate another
rank--$r'$ root system, where $r'={\rm min}(p,q)$ is the rank of the
symmetric space. For some symmetric spaces, it is the same as the root
system inherited from the complex extension algebra (see Table~1 for a
list of the restricted root systems), but this need not be the case.
For example, the restricted root system is, in the case of
$SU(p,q;C)/(SU(p)\otimes SU(q)\otimes U(1))$, $BC_{r'}$. When it is
the same and when it is different, as well as why the rank can change,
will be obvious from the example we will give below.

In general the restricted root system will be different from the
original, inherited root system if the Cartan subalgebra lies in ${\bf
  K}$. The procedure to find the restricted root system is then to
define an {\it alternative Cartan subalgebra} that lies
partly (or entirely) in ${\bf P}$ (or $i{\bf P}$).

To achieve this, we first look for a different representation of the
original Cartan subalgebra, that gives the same root lattice as the
original one (i.e., $A_{n-1}$ for the ${\bf SU(p,q;C)}$ algebra). In
general, this root lattice is an automorphism of the original root
lattice of the same kind, obtained by a permutation of the roots.
Unless we find this new representation, we will not be able to find a
new, alternative Cartan subalgebra that lies partly in the subspace
${\bf P}$.

Once this has been done, we take a maximal abelian subalgebra of ${\bf
  P}$ (the number of generators in it will be equal to the rank $r'$
of the symmetric space $G/K$ or $G^*/K$) and find the generators in
${\bf K}$ that commute with it. These generators will be among the
ones that are in the new representation of the original Cartan
subalgebra.  These commuting generators now form our new,
alternative Cartan subalgebra that lies partly in ${\bf P}$,
partly in ${\bf K}$.  Let's call it ${\bf A_0}$.

The new root system is defined with respect to the part of the maximal
abelian subalgebra that lies in ${\bf P}$. Therefore its rank is
normally smaller than the rank of the root system inherited from the
complex extension. We can define raising and lowering operators
$E'_\alpha$ in the {\it whole} algebra ${\bf G}$ that satisfy

\beq
[X'_i,E'_\alpha]=\alpha'_i E'_\alpha \ \ \ \ \ \ \ (X'_i\in {\bf A_0}\cap
{\bf P})
\eeq

The roots $\alpha'_i$ define the restricted root system.  

{\bf Example:} Let's now look at a specific example. We will start
with the by now familiar algebra ${\bf SU(3,C)}$. As before, we use
the convention of regarding the $T_i$'s as the generators, without the
$i$ in front (recall that the algebra consists of elements of the form
$\sum_at^aX_a=i\sum_at^aT_a$; cf. the footnote in conjuction with
eq.~(\ref{eq:Gell-Mann})).  In subsection \ref{sec-rootsp} we
explicitly constructed its root lattice $A_2$. Let's write down the
generators again:

\beq
\label{eq:Gell-Mann'}
\begin{array}{l}
T_1=\frac{1}{2}\left(\begin{array}{ccc} 0 & 1 & 0 \\
                                   1 & 0 & 0 \\
                                   0 & 0 & 0 \end{array}\right), \ \ \ \
T_2=\frac{1}{2}\left(\begin{array}{ccc} 0 & -i & 0 \\
                                   i & 0 & 0 \\
                                   0 & 0 & 0 \end{array}\right), \ \ \ \
T_3=\frac{1}{2}\left(\begin{array}{ccc} 1 & 0 & 0 \\
                                   0 & -1 & 0 \\
                                   0 & 0 & 0 \end{array}\right), \nonumber \\
\\
T_4=\frac{1}{2}\left(\begin{array}{ccc} 0 & 0 & 1 \\
                                   0 & 0 & 0 \\
                                   1 & 0 & 0 \end{array}\right), \ \ \ \  
T_5=\frac{1}{2}\left(\begin{array}{ccc} 0 & 0 & -i \\
                                   0 & 0 & 0 \\
                                   i & 0 & 0 \end{array}\right), \ \ \ \
T_6=\frac{1}{2}\left(\begin{array}{ccc} 0 & 0 & 0 \\
                                   0 & 0 & 1 \\
                                   0 & 1 & 0 \end{array}\right), \nonumber \\ 
\\ 
T_7=\frac{1}{2}\left(\begin{array}{ccc} 0 & 0 & 0 \\
                                   0 & 0 & -i \\
                                   0 & i & 0 \end{array}\right), \ \ \ \
T_8=\frac{1}{2\sqrt{3}}\left(\begin{array}{ccc} 1 & 0 & 0 \\
                                                0 & 1 & 0 \\
                                                0 & 0 & -2 
                                   \end{array}\right) \end{array}
\\ \nonumber
\eeq

The splitting of the ${\bf SU(3,C)}$ algebra in terms of the subspaces
${\bf K}$ and ${\bf P}$ was given in eq.~(\ref{eq:SU3}):

\beq
{\bf K}=\{iT_2,iT_5,iT_7\} ,\ \ \ \ \ \ \
{\bf P}=\{iT_1,iT_3,iT_4,iT_6,iT_8\}
\eeq

The Cartan subalgebra is $\{ iT_3,iT_8\} $.
The raising and lowering operators were given in (\ref{eq:raislowSU3})
in terms of $T_i$:

\beq
\label{eq:raislowSU3'}
\begin{array}{l}
E_{\pm(1,0)}=\frac{1}{\sqrt{2}}(T_1\pm iT_2) \\ \\
E_{\pm(\frac{1}{2},\frac{\sqrt{3}}{2})}=\frac{1}{\sqrt{2}}(T_4\pm iT_5) \\ \\
E_{\pm(-\frac{1}{2},\frac{\sqrt{3}}{2})}=\frac{1}{\sqrt{2}}(T_6\pm iT_7) 
\end{array} 
\eeq

Now let us construct the Cartan decomposition of ${\bf G'^*}={\bf
  K'}\oplus i{\bf P'}= {\bf SU(2,1;C)}$. We know from paragraph
\ref{sec-realforms2} that ${\bf K'}$ and ${\bf P'}$ are given by
matrices of the form

\beq
\left( \begin{array}{cc} A & 0 \\
                          0 & C \end{array}\right) \in {\bf K'},
\ \ \ \ \ \ \ 
\left( \begin{array}{cc} 0 & B \\
                          -B^\dagger & 0 \end{array}\right) \in {\bf P'}
\eeq

where $A$ and $C$ are antihermitean and ${\rm tr}A+{\rm tr}C=0$.
Combining the generators to form this kind of block--structures (or
alternatively, using the involution $\sigma_2=I_{2,1}$) we need to
take linear combinations of the $X_i$'s, with real coefficients, and
we then see that the subspaces ${\bf K'}$ and $i{\bf P'}$ are spanned
by

\beq 
\begin{array}{l}
{\bf K'}=\left\{ 
\frac{i}{2}\left(\begin{array}{ccc}  0 & 1 &   \\
                                     1 & 0 &   \\
                                       &   & 0 \end{array}\right),
\frac{1}{2}\left(\begin{array}{ccc}  0 & 1 &   \\
                                    -1 & 0 &   \\
                                       &   & 0 \end{array}\right),
\frac{i}{2}\left(\begin{array}{ccc}  1 & 0  &   \\
                                     0 &-1 &   \\
                                       &   & 0 \end{array}\right),
\frac{i}{2\sqrt{3}}\left(\begin{array}{ccc} 1  & 0 &   \\
                                            0  & 1 &   \\
                                               &   & -2 \end{array}\right)
\right\} \\ \\
=\{ iT_1,iT_2,iT_3,iT_8\}\\  \\ 
i{\bf P'}=\left\{ 
\frac{1}{2}\left(\begin{array}{ccc}    &   & 1  \\
                                       &   & 0  \\
                                     1 & 0 &    \end{array}\right),
\frac{i}{2}\left(\begin{array}{ccc}    &   & -1  \\
                                       &   & 0  \\
                                    1  & 0 &    \end{array}\right),
\frac{1}{2}\left(\begin{array}{ccc}    &   & 0  \\
                                       &   & 1  \\
                                     0 & 1 &    \end{array}\right),
\frac{i}{2}\left(\begin{array}{ccc}    &   & 0  \\
                                       &   & -1  \\
                                    0  & 1 &    \end{array}\right)
\right\}\\  \\
=\{ T_4,T_5,T_6,T_7\}\end{array}
\eeq

where the block--structure is evidenced by leaving blank the remaining
zero entries. ${\bf K'}$ spans the algebra of the symmetric subgroup
$SU(2)\otimes U(1)$ and $i{\bf P'}$ spans the complementary subspace
corresponding to the symmetric space $SU(2,1)/(SU(2)\otimes U(1))$.
$i{\bf P'}$ is spanned by matrices of the form

\beq
\left(\begin{array}{cc} 0 & \tilde{B}\\ \tilde{B}^\dagger & 0\end{array}\right)
\eeq

We see that the Cartan subalgebra $i{\bf H_0}=\{iT_3,iT_8\}$ lies
entirely in ${\bf K'}$.  It is easy to see that by using the
alternative representation

\beq
T'_3=\frac{1}{2}\left(\begin{array}{ccc}  1 &   &   \\
                                            &0  &   \\
                                            &   & -1 \end{array}\right),
\ \ \ \ \ \ \ 
T'_8=\frac{1}{2\sqrt{3}}\left(\begin{array}{ccc}  1 &   &   \\
                                                    &-2 &   \\
                                                    &   & 1 \end{array}\right)
\eeq

of the Cartan subalgebra (note that this is a valid representation of
${\bf SU(3,C)}$ generators) while the other $T_i$'s are unchanged, we
still get the same root lattice $A_2$. The eigenvectors under the
adjoint representation, the $E_\alpha $'s, are still given by
eq.~(\ref{eq:raislowSU3'}). However, their eigenvalues (roots) are
permuted under the new adjoint representation of the 
Cartan subalgebra, so that they no longer correspond to the root
subscripts in (\ref{eq:raislowSU3'}).

Now we choose the alternative Cartan subalgebra to consist of
the generators $T_4$, $T'_8$: 

\beq
{\bf A_0}=\{T_4,T'_8\},\ \ \ [T_4,T'_8]=0, \ \ \ \ \ \ \ iT_4\in {\bf P'},
\ iT'_8\in {\bf K'}
\eeq

(Note that unless we first take a new representation of the original
Cartan subalgebra, we are not able to find the alternative Cartan
subalgebra that lies partly in ${\bf P'}$.) The restricted root system
is now about to be revealed.  We define raising and lowering operators
$E'_\alpha$ in the whole algebra according to

\beq
E'_{\pm 1}\sim (T_5\pm iT_3)\ \ \ \  
E'_{\pm \frac{1}{2}}\sim (T_6\pm iT_2)\ \ \ \  
\tilde{E}'_{\pm \frac{1}{2}}\sim (T_7\pm iT_1) 
\eeq

The $\pm\alpha $ subscripts are the eigenvalues of $T_4\in i{\bf
  P'}$ in the adjoint representation:

\bea
\label{eq:BC_1}
[T_4,E'_{\pm 1}]=\pm E'_{\pm 1},\ \ \ [T_4,E'_{\pm \frac{1}{2}}]=\pm
\frac{1}{2} E'_{\pm \frac{1}{2}},\ \ \ [T_4,\tilde{E}'_{\pm \frac{1}{2}}]=
\pm \frac{1}{2}\tilde{E}'_{\pm \frac{1}{2}}
\eea

These roots form a one--dimensional root system of type $BC_1$. We see
that the multiplicity of the long roots is $1$ and the multiplicity of
the short roots is $2=2(p-q)$. This result is general (cf. Table~1).
If we had ordinary roots, their multiplicity would be $2$, but for
this low--dimensional group we can have only $3$ pairs of roots. Note
that we can rescale the lengths of all the roots together by rescaling
the operator $T_4$ in (\ref{eq:BC_1}), but their characters as long
and short roots can not change. The root system $BC_1$ is with respect
to the part of the Cartan subalgebra lying in $i{\bf P'}$ only, thus
it is called restricted.

\begin{table}[ht]
{\bf Table~1.}\ Irreducible symmetric spaces of positive and negative
curvature originating in simple Lie groups. In the third and
fourth columns the symmetric spaces $G/K$ and $G^*/K$ are listed for
all the entries except the simple Lie groups themselves, for which the
symmetric spaces $G$ and $G^C/G$ are listed.  Note that there are also
zero curvature spaces corresponding to non--semisimple groups and
isomorphic to the subspace ${\bf P}$ of the algebra, when ${\bf P}$ is
an abelian invariant subalgebra. These are not listed in the table,
but can be constructed as explained in subsection \ref{sec-curv}. The
root multiplicities listed pertain to the {\it restricted} root
systems of the pairs of dual symmetric spaces with positive and negative 
curvature.  
\vskip5mm

\hskip-.8cm
\begin{tabular}{|l|l|l|l|l|l|l|l|}
\hline
$\begin{array}{c}Root\\ space\end{array}$&
$\begin{array}{c}Restricted\\ root\ space\end{array}$ & $\begin{array}{c}Cartan\\ class \end{array}$ & $G/K\ (G)$ & $G^*/K\ (G^C/G)$ & $m_o$ & $m_l$ & $m_s$ \\
\hline

$A_{N-1}$ & $A_{N-1}$     & A    & $SU(N)$                 & $\frac{SL(N,C)}{SU(N)}$ & 2 & 0 & 0 \\   
          & $A_{N-1}$     & AI   & $\frac{SU(N)}{SO(N)}$   & $\frac{SL(N,R)}{SO(N)}$ & 1 & 0 & 0 \\  
          & $A_{N-1}$     & AII  & $\frac{SU(2N)}{USp(2N)}$ & $\frac{SU^*(2N)}{USp(2N)}$ & 4 & 0 & 0  \\
          &\hskip-2mm $\begin{array}{l} BC_q\ {\scriptstyle (p>q)} \\ C_q\  {\scriptstyle (p=q)} \end{array}$ 
                          & AIII & $\frac{SU(p+q)}{SU(p)\times SU(q)\times U(1)}$ & $\frac{SU(p,q)}{SU(p)\times SU(q)\times U(1)}$ & 2 & 1 & $2(p-q)$ \\
\hline

$B_N$ &$B_N$          & B    & $SO(2N+1) $               & $\frac{SO(2N+1,C)}{SO(2N+1)}$ & 2 & 0 & 2 \\

\hline

$C_N$  & $C_N$    & C    & $USp(2N)$                               &$\frac{Sp(2N,C)}{USp(2N)}$ & 2 & 2 & 0 \\
       & $C_N$    & CI   & $\frac{USp(2N)}{SU(N)\times U(1)}$      & $\frac{Sp(2N,R)}{SU(N)\times U(1)}$& 1 & 1 & 0  \\
       &\hskip-2mm $\begin{array}{l} BC_q\  {\scriptstyle (p>q)} \\  C_q\  {\scriptstyle (p=q)} \end{array}$ 
                  & CII & $\frac{USp(2p+2q)}{USp(2p)\times USp(2q)}$ & $\frac{USp(2p,2q)}{USp(2p)\times USp(2q)}$& 4 & 3 & $4(p-q)$ \\
\hline

$D_N$& $D_N$     & D          &   $SO(2N)$                            & $\frac{SO(2N,C)}{SO(2N)}$ & 2 & 0 & 0 \\
     &$C_N$     & DIII-even  & $\frac{SO(4N)}{SU(2N)\times U(1)}$     & $\frac{SO^*(4N)}{SU(2N)\times U(1)}$ & 4 & 1 & 0 \\
     &$BC_N$    & DIII-odd   & $\frac{SO(4N+2)}{SU(2N+1)\times U(1)}$ & $\frac{SO^*(4N+2)}{SU(2N+1)\times U(1)}$& 4 & 1 & 4 \\

\hline
\hskip-2mm $\begin{array}{l} B_N\ {\scriptstyle (p+q=2N+1)}\\ D_N\ {\scriptstyle (p+q=2N)}\end{array}$  
          &\hskip-2mm $\begin{array}{l} B_q\ {\scriptstyle (p>q)}\\ D_q\ {\scriptstyle (p=q)} \end{array}$
            & BDI & $\frac{SO(p+q)}{SO(p)\times SO(q)}$   & $\frac{SO(p,q)}{SO(p)\times SO(q)}$ & 1 & 0 & $p-q$ \\

\hline
\end{tabular}
\end{table}
  
\subsection{Real forms of symmetric spaces}
\label{sec-realformsSS}

Involutive automorphisms were used to 
split the algebra ${\bf G}$ into orthogonal subspaces
to obtain the real forms ${\bf G}$, ${\bf G^*}$, ${\bf G'^*}$... 
of a complex extension algebra ${\bf G^C}$. By re--applying the same 
involutive automorphisms to the spaces ${\bf K}$,
${\bf P}$, and $i{\bf P}$, these spaces with a definite metric tensor can in 
turn be split into subspaces with eigenvalue $+1$ and $-1$ under this 
new involutive automorphism $\tau $. Thus,

\beq
\begin{array}{l}
\sigma : {\bf G}\to {\bf K}\oplus {\bf P},\\
\tau :{\bf K}\to {\bf K_1}\oplus {\bf K_2},\\
\tau :{\bf P}\to {\bf P_1}\oplus {\bf P_2},\\
\tau :i{\bf P}\to i{\bf P_1}\oplus i{\bf P_2},\end{array}\ \ 
\begin{array}{l}
{\bf G^*}={\bf K}\oplus i{\bf P}\\
{\bf H}={\bf K_1}\oplus i{\bf K_2}\\
{\bf M}={\bf P_1}\oplus i{\bf P_2}\\ 
i{\bf M}=i{\bf P_1}\oplus {\bf P_2} \end{array}
\eeq

As we already know, $K$ is a compact subgroup, and $\exp ({\bf P})$
and $\exp (i{\bf P})$ define symmetric spaces with a {\it definite}
metric (Riemannian spaces). In the same way, $H$ is a non--compact
subgroup, $\exp ({\bf M})$ and its dual space $\exp (i{\bf M})$ define
symmetric spaces with an {\it indefinite} metric. These are
pseudo--Riemannian symmetric coset spaces of a non--compact group by a
maximal non--compact subgroup\footnote{Note that not all the theorems
governing symmetric spaces corresponding to maximal compact subgroups
apply to the case at hand. A prime example is the decomposition
involving radial coordinates in subsection \ref{sec-radial}. We will
not discuss the symmetric spaces involving maximal non--compact
subgroups in any detail in this paper.}.  The original algebra ${\bf G}$
is thereby split into four components ${\bf K_1}$, ${\bf K_2}$, ${\bf
P_1}$, ${\bf P_2}$, depending on their eigenvalues ($++,+-,-+,--$)
under the two successive automorphisms $\sigma $, $\tau $. By applying
all the possible $\sigma $'s and all the possible $\tau $'s, or by
replacing either $\sigma $ or $\tau $ by the involutive automorphism
$\sigma \tau = \tau \sigma $, we obtain all the possible {\it real
forms of the symmetric spaces} associated with the compact algebra
${\bf G}$.

{\bf Example:} The complex algebra ${\bf SO(3,C)}$ has a root system
of type $B_n$. Its compact real form is ${\bf SO(3,R)}$, and its only
non--compact real form is ${\bf SO(p,q;R)}\simeq {\bf SO(q,p;R)}$
($p+q=3$), obtained by applying the involution $\sigma $=$I_{p,q}$
($I_{q,p}$) to ${\bf SO(3,R)}$. In paragraph \ref{sec-realforms2} we
constructed two Riemannian symmetric spaces associated with the
algebra ${\bf SO(3)}$, the sphere $SO(3)/SO(2)$ and the
double--sheeted hyperboloid $SO(2,1)/SO(2)$.  The Killing form has a
definite opposite sign for the two spaces.

The single--sheeted hyperboloid, described by the equation
$-x^2+y^2+z^2=1$ in ${\bf R}^3$, corresponds to the pseudo--Riemannian
symmetric space $SO(2,1)/SO(1,1)$ associated with the same algebra.
It is obtained by applying two consecutive involutive automorphisms
$\sigma $=$I_{2,1}$, $\tau $=$I_{1,2}$ to the algebra ${\bf G}={\bf
  SO(3,R)}$. Like in eq.~(\ref{eq:SO(2,1)}), $I_{2,1}$ and the Weyl
unitary trick transforms ${\bf G}$ into ${\bf G^*}$. Let's now apply
$I_{1,2}$ to ${\bf G^*}$:

\beq \begin{array}{l}
I_{1,2}XI_{1,2}^{-1}=\left(\begin{array}{ccc} 1 &    &    \\
                                           &  -1 &    \\
                                           &    & -1 \end{array}\right)
\frac{1}{2}\left(\begin{array}{ccc}     & t^3 & it^2\\
                   -t^3 &     & it^1\\
                   -it^2 & -it^1& \end{array}\right)
\left(\begin{array}{ccc} 1 &    &    \\
                                           &  -1 &    \\
                                           &    & -1 \end{array}\right) \\
\\
=\frac{1}{2}\left(\begin{array}{ccc}     & -t^3 & -it^2\\
                   t^3 &     & it^1\\
                   it^2 & -it^1& \end{array}\right)=
\frac{1}{2}\left(\begin{array}{ccc}     & -t^3 & \\
                   t^3 &     & \\
                    & & \end{array}\right)\oplus
\frac{1}{2}\left(\begin{array}{ccc}     &  & -it^2\\
                    &     & it^1\\
                   it^2 & -it^1& \end{array}\right)\\
\\
=({\bf K_1}\oplus {\bf K_2})\oplus i({\bf P_1}\oplus {\bf P_2})
\end{array}
\eeq

where in this example, ${\bf K_1}$ is empty. The spaces ${\bf K_1}$, 
${\bf K_2}$, ${\bf P_1}$, ${\bf P_2}$ consist of the generators in ${\bf G}$
with the following combinations  of eigenvalues under the two successive 
involutions, $\sigma \tau$:

\beq
{\bf K_1}: ++\ \ \ \ \ {\bf K_2}: +-\ \ \ \ \ {\bf P_1}: -+\ \ \ \ \ {\bf P_2}: --
\eeq

Thus we see that  ${\bf K_1}$ is empty and the others are spanned by 

\beq
{\bf K_2}=\left\{\frac{1}{2}\left(\begin{array}{ccc} & 1 & \\
                                                   -1 & & \\
                                          & & \end{array}\right)\right\},\ \ \ 
{\bf P_1}=\left\{\frac{1}{2}\left(\begin{array}{ccc} &  & \\
                                                    & &1 \\
                                        &-1 & \end{array}\right)\right\},\ \ \ 
{\bf P_2}=\left\{\frac{1}{2}\left(\begin{array}{ccc} &  & 1\\
                                                    & & \\
                                         -1 & & \end{array}\right)\right\}
\eeq

The new symmetric space is obtained by doing the Weyl unitary trick on 
the split spaces (${\bf K_1}\oplus {\bf K_2}$) and (${\bf P_1}\oplus 
{\bf P_2}$):

\beq \begin{array}{l}
{\bf H}={\bf K_1}\oplus i{\bf K_2}
=\frac{1}{2}\left(\begin{array}{ccc} & it^3 & \\
                               -it^3 & & \\
                                     & & \end{array}\right)\\ \\ 
{\bf M}={\bf P_1}\oplus i{\bf P_2}=
\frac{1}{2}\left(\begin{array}{ccc} &  & it^2\\
                                    & & t^1\\
                              -it^2 & -t^1 & \end{array}\right)
\end{array}
\eeq

The second involution $\tau $ (plus the Weyl trick) gives rise to a
non--compact subgroup $H$=$SO(1,1)$ and to the symmetric space $M\sim 
{\rm exp}{\bf M}$ and its dual $M^*\sim {\rm exp}(i{\bf M})$. The coset  
$M\sim SO(2,1)/SO(1,1)$ is represented by 

\beq
{\rm exp}{\bf M}=\left(\begin{array}{ccc} . & .  & ix\\
                         . & .  & y\\
                         -ix &  -y & z   \end{array}\right);\ \ \ \ 
(ix)^2+y^2+z^2=1
\eeq

The real forms of the simple Lie groups do not include all the
possible Riemannian symmetric coset spaces. For example, the compact Lie group 
$G$ is itself such a space, and so is its dual $G^C/G$ (here the algebra
${\bf G^C}={\bf G^*}\oplus i{\bf G^*}$ is the complex extension of
all the real forms ${\bf G^*}$). By starting with a compact algebra 
${\bf G}$ and applying to it all the combinations of the two involutive 
automorphisms $\sigma $, $\tau $, we construct, in the way just described,
all the remaining pseudo--Riemannian
symmetric spaces associated to the corresponding root system.
A complete list of these spaces can be found in Table~9.7 of 
reference~\cite{Gilmore}. 

Note that all the properties of the Lie algebra ${\bf G}$ (Killing form, 
rank, and so on) can be transferred to the vector 
subspaces ${\bf P}$, $i{\bf P}$ \cite{Gilmore}.
The only difference is that the subspaces are not closed under commutation.

\section{Operators on symmetric spaces}
\label{sec-Operators}

The differential operator uniquely determined by the simplest Casimir
operator on a symmetric space (and especially its radial part) plays
an important role both in mathematics and in the physical applications
of symmetric spaces.  Its eigenfunctions provide a complete basis for
the expansion of an arbitrary square--integrable function on the
symmetric space, and are therefore important in their own right.
Their importance in the applications to be discussed in \cite{CasMag}
is evident when considering that the radial part of the
Laplace--Beltrami operator on an underlying symmetric space determines
the dynamics of the transfer matrix eigenvalues of the DMPK
equation in the theoretical description of quantum wires, and maps
onto the Hamiltonians of integrable Calogero--Sutherland models.  Here
we will define some concepts related to the Laplace--Beltrami operator
and discuss its eigenfunctions.

\subsection{Casimir operators}
\label{sec-Casi}

Let ${\bf G}$ be a semisimple rank--$r$ Lie algebra. A {\it Casimir operator} 
(invariant operator) $C_k$ ($k=1,...,r$)
associated with the algebra ${\bf G}$ is a homogeneous polynomial 
operator that satisfies

\beq
[C_k,X_i]=0
\eeq

for all $X_i\in {\bf G}$. The simplest (quadratic) Casimir
operator associated to the adjoint representation of the 
algebra ${\bf G}$ is given by 

\beq
\label{eq:C}
C=g^{ij}X_iX_j
\eeq

where $g^{ij}$ is the inverse of the metric tensor defined in (\ref{eq:metric})
and the generators $X_i$ are in the adjoint representation.
More generally, it can be defined for any representation  $\rho $ of ${\bf G}$ 
by

\beq
C_\rho =g_\rho^{ij}\rho (X_i)\rho (X_j)
\eeq

where $g_\rho^{ij}$ is the inverse of the metric (\ref{eq:g_rho})
for the representation $\rho$ (cf.
subsection {\ref{sec-metric}).  The Casimir operators lie in the
enveloping algebra obtained by embedding ${\bf G}$ in the associative 
algebra defined by the relations

\beq
\label{eq:asso}
X(YZ) = (XY)Z  \ \ \ \ \ \ \ [X,Y] = XY-YX 
\eeq

(note that in general, $XY$ makes no sense in the algebra ${\bf G}$).

The number of functionally independent Casimir operators is equal to
the rank $r$ of the group.  Other Casimir operators can be formed by
taking polynomials of the independent Casimir operators $C_k$
($k=1,...,r$).  Since the Casimir operators commute with all the
elements in ${\bf G}$, they make up the center of the associative
algebra (\ref{eq:asso}).

Note that Casimir operators are defined for {\it semisimple} algebras,
where the metric tensor has an inverse. This does not prevent one from
finding operators that commute with all the generators of
non--semisimple algebras.  For example, for the euclidean group $E_3$
of rotations $\{J_1,J_2,J_3\} $ and translations $\{P_1,P_2,P_3\} $,
${\bf P}^2=\sum P_iP_i$ and ${\bf P\cdot J}=\sum P_iJ_i$ commute with
all the generators. Also these analogous operators commuting with all
the group elements associated to semisimple groups are often referred
to as Casimir operators.

All the independent Casimir operators of the algebra ${\bf G}$ can be
obtained as follows.  Suppose $\rho $ is an $n$--dimensional
representation of the rank--$r$ Lie algebra ${\bf G}$.  The {\it
  secular equation} for the algebra ${\bf G}$ is defined as the
eigenvalue equation

\beq
{\rm det}\left(\sum_{i=1}^{{\rm dim}{\bf G}} 
t^i\rho(X_i) -\lambda I_n\right)=\sum_{k=0}^n (-\lambda)^{n-k}\varphi_k(t^i)=0
\eeq

where the $\varphi_k(t^i)$ are functions of the real coordinates
$t^i$. In general, they will not all be functionally independent (for
example, $\varphi_0(t^i)$ is a constant).  There will be $r$
functionally independent coefficients $\varphi_k(t^i)$ multiplying the
powers of $-\lambda $ \cite{Gilmore}. When writing down the secular
equation, it is easiest to take a low--dimensional representation. By
making the substitution $t^i\to X_i$ in the functionally independent
coefficients, they become the functionally independent Casimir
operators of the algebra ${\bf G}$:

\beq
\label{eq:substitutionC}
\varphi_k(t^i)\begin{array}{c}{\scriptstyle t^i\to X_i}\\ 
\longrightarrow \\ {} \end{array} C_l(X_i)
\eeq

{\bf Example:} The generators $L_1$, $L_2$, $L_3$ of the ${\bf SO(3)}$
algebra were given explicitly in the adjoint representation in
equations (\ref{eq:L1L2}), (\ref{eq:L3}) in subsection
\ref{sec-cosets}. The secular equation for this algebra is then

\beq 
{\rm det}\left({\bf t\cdot L}-\lambda I_3\right)=
\left|\begin{array}{ccc} -\lambda & t^3/2 & t^2/2 \\
                                   -t^3/2 & -\lambda & t^1/2 \\
                                   -t^2/2 & -t^1/2 & -\lambda \end{array}\right|
=(-\lambda)^3+(-\lambda)\frac{1}{4}{\bf t}^2=0
\eeq

The equation has one functionally independent coefficient, which is
proportional to the trace of the matrix $({\bf t\cdot L})^2$. It equals
$\varphi_1({\bf t}) =\frac{1}{4}{\bf t}^2$. The rank of $SO(3)$ is $1$
and the only Casimir operator is

\beq
C_1\sim {\bf L}^2=L_1^2+L_2^2+L_3^2
\eeq

It is obtained by the substitution $t^i\to L_i$ in $\varphi_1({\bf t})$.
The Casimir operator can also be obtained from eq.~(\ref{eq:C}) by using
the metric $g_{ij}=-\frac{1}{2}\delta_{ij}$ for ${\bf SO(3)}$ given in
the example in subsection \ref{sec-metric}. 
We know from elementary quantum mechanics that 

\beq
\label{eq:L^2commrel}
[{\bf L}^2,L_1]=[{\bf L}^2,L_2]=[{\bf L}^2,L_3]=0
\eeq

is an immediate consequence of the commutation relations, so we see
that this operator indeed commutes with all the generators.  Even
though the commutation relations are not the same in polar
coordinates or after a general coordinate transformation, 
(\ref{eq:L^2commrel}) will nevertheless be true.

{\bf Example:} $SU(3)$ is a rank--2 group and therefore its characteristic
equation will have two independent coefficients. If we denote a general 
${\bf SU(3)}$ matrix $(a_{ij})$ we get the characteristic equation

\beq
\begin{array}{l}
{\rm det}\left( \begin{array}{ccc} a_{11}-\lambda & a_{12} & a_{13} \\
                                   a_{21} & a_{22}-\lambda & a_{23} \\
                                   a_{31} & a_{32} & a_{33}-\lambda 
\end{array}\right)
=(-\lambda )^3 + (-\lambda )^2 (a_{11}+a_{22}+a_{33})\\ \\ +
(-\lambda )(a_{11}a_{22}+a_{22}a_{33}+a_{33}a_{11}-a_{12}a_{21}-a_{23}a_{32}-
a_{31}a_{13})\\ \\ +(a_{11}(a_{22}a_{33}-a_{23}a_{32})+a_{12}(a_{23}a_{31}-
a_{21}a_{33})+a_{13}(a_{32}a_{21}-a_{31}a_{22}))=0
\end{array}
\eeq

The term proportional to $(-\lambda )^2$ vanishes, because the trace
of any matrix in the ${\bf SU(3)}$ algebra is zero. The two
independent coefficients are then $\varphi_2(a_{ij})$ and
$\varphi_3(a_{ij})$. Substituting the values in terms of the
coordinates $t^i$ of the algebra $\sum_i t^iT_i$ for the $a_{ij}$ (for
example, $a_{11}=t^3+\frac{1}{\sqrt{3}}t^8$, $a_{12}=t^1+it^2$, etc.),
we see that the expression for $\varphi_2(t^i)$ becomes

\beq
\varphi_2(t^i)=\sum_{i=1}^8 (t^i)^2
\eeq

and therefore the substitution (\ref{eq:substitutionC}) gives the 
first Casimir operator 

\beq
C_1=H_1^2+H_2^2+\sum_\alpha 
(E_\alpha E_{-\alpha }+E_{-\alpha }E_\alpha )=
\sum_{i=1}^8 T_i^2
\eeq

as expected. Making the same substitution in $\varphi_3(t^i)$ gives the second
Casimir operator for $SU(3)$, which has a more complicated form.

\subsection{Laplace operators}
\label{sec-Laplaceop}

The Casimir operators can be expressed as differential
operators in the local coordinates on the symmetric space. This is due
to the fact that each infinitesimal generator $X_\alpha \in {\bf G}$ is a
contravariant vector field on the group manifold. An element in the Lie algebra
can be written 

\beq
X=\sum_\alpha X^\alpha(x) X_\alpha \equiv \sum_\alpha X^\alpha(x) \frac{\partial }{\partial x^\alpha }
\eeq

where $x^\alpha $ are local coordinates \cite{Helgason,SattW} (for
example, $L_1=({\bf r \times P})_1= x^2\partial_3-x^3\partial_2$).
That the generators transform as lower index objects follows from the
commutation relations.

{\bf Example:} As an example we take the group $SO(3)$.  Under a
rotation $R=R(t^1,t^2,t^3)={\rm exp}(\sum t^kL_k)$, the vector ${\bf
  x}=x^i\hat{e}_i\ \in {\bf R}^3$ transforms as

\beq
{\bf x} {\buildrel{\scriptstyle R} \over \longrightarrow } {\bf x'}=x'^i\hat{e}'_i
\eeq

where the transformation laws for the components and the natural basis vectors
are 

\beq
x'^i=R^i_{\ j}x^j,\ \ \ \ \hat{e}'_i=\hat{e}_jR^j_{\ i}
\eeq

and $R^{-1}=R^T$.  The one--parameter subgroups of $SO(3)$ are
rotations

\beq
R(t^n)={\rm exp}(t^nL_n), \ \ \ \ \ \ \ (n=1,2,3) 
\eeq

(no summation) where $L_n$ are $SO(3)$ generators. It is easy to show
using the commutation relations for $L_n$ (given after
eq.~(\ref{eq:fixedNP})) that under infinitesimal rotations the $L_n$
transform like the lower index objects $\hat{e}_i$:

\beq
RL_iR^{-1}=L_jR^j_{\ i}
\eeq

Expressed in local coordinates as differential operators, the Casimirs
are called {\it Laplace operators}. In analogy with the Laplacian in
${\bf R^n}$,

\beq
{\bf P}^2=\Delta=\sum_{i=1}^n\frac{\partial^2}{\partial {x^i}^2}
\eeq

which is is invariant under the group $E_n$ of rigid motions
(isometries) of ${\bf R^n}$, the Laplace operators on
(pseudo--)Riemannian manifolds are invariant under the group of
isometries of the manifold.  The isometry group of the symmetric space
$P\simeq G/K$ is $G$, since $G$ acts transitively on this space and
preserves the metric, so the Laplace operators are invariant under the
group operations $g\in G$.
 
The number of independent Laplace operators on a Riemannian symmetric
coset space is equal to the rank of the space. As we defined in paragraph
\ref{sec-action}, the rank of a symmetric space is the maximal number
of mutually commuting generators $H_i$ in the subspace ${\bf P}$ (cf. also
subsection \ref{sec-restricted}).  If $X_\alpha$, $X_\beta,... \in
{\bf K}$ and $X_i$, $X_j,... \in {\bf P}$, it is also equal to the
number of functionally independent solutions to the equation

\beq
{\rm det}\left(\sum_{k=1}^{{\rm dim}{\bf P}} 
t^k\rho(X_k) -\lambda I_n\right)=\sum_{l=0}^n (-\lambda)^{n-l}\varphi_l(t^k)=0
\eeq

where now in the determinant we sum over all $X_k\in {\bf P}$.  This
is equivalent to setting the coordinates $t^\gamma $ for all the
$X_\gamma\in {\bf K}$ equal to zero in the secular equation. In the
example in the preceding paragraph, the rank of the symmetric space
$SO(3)/SO(2)$ (the 2--sphere) is $1$, which in this case is also the
rank of the group $SO(3)$.

The {\it Laplace--Beltrami operator} on a symmetric space is the
  special second order Laplace operator defined (when acting on a
  function (0--form) $f$) as

\beq
\label{eq:L-B-op}
\Delta_Bf=g^{ij}D_iD_jf=g^{ij}(\partial_i\partial_j-\Gamma^k_{ij}\partial_k)f
=\frac{1}{\sqrt{|g|}}\frac{\partial }{\partial x^i} g^{ij}
\sqrt{|g|}\frac{\partial }{\partial x^j}f,\ \ \ \ \ \ \ 
g\equiv {\rm det}g_{ij}
\eeq

Here $D_i$ denotes the covariant derivative on the symmetric space. It
is defined in the usual way \cite{Helgason,FosNigh,3w}, for example
it acts on the components $x^j$ of a contravariant vector field in the
following way:
 
\beq
D_ix^j=\partial_ix^j+\Gamma^j_{ki}x^k
\eeq

where $\Gamma^j_{ki}$ are Christoffel symbols (connection coefficients).
The last term represents the change in $x^j$
due to the curvature of the space. We remind the reader that on a Riemannian 
manifold, the $\Gamma^j_{ki}$ are expressible in terms of the metric tensor,
hence the formula in eq.~(\ref{eq:L-B-op}).

{\bf Example:} Let's calculate the Laplace--Beltrami operator on the
symmetric space $SO(3)/SO(2)$ in polar coordinates using
(\ref{eq:L-B-op}) and the metric at the point $(\theta,\phi)$ given in
the second example of subsection \ref{sec-metric}:

\beq
g_{ij}=\left(\begin{array}{cc} 1 & 0 \\
                       0 & {\rm sin}^2\theta \end{array}\right),\ \ \ \ \ \ \  
g^{ij}=\left(\begin{array}{cc} 1 & 0 \\
                       0 & {\rm sin}^{-2}\theta \end{array}\right) 
\eeq

Substituting in the formula and computing derivatives we obtain the 
Laplace--Beltrami operator on the sphere of radius $1$:

\beq
\label{eq:Delta_on_sphere}
\Delta_B=\partial_\theta^2+{\rm cot}\theta\, \partial_\theta+
{\rm sin}^{-2}\theta \, \partial_\phi^2
\eeq

Of course this operator is exactly ${\bf L}^2$. We can check this by 
computing $L_x= y\partial_z-z\partial_y$, $L_y= z\partial_x-x\partial_z$,
and $L_z= x\partial_y-y\partial_x$ in spherical coordinates (setting $r=1$)
and then forming the operator $L_x^2+L_y^2+L_z^2$, remembering that all the
operators have to act also on anything coming after the expression for each 
$L_i^2$. We find that  ${\bf L}^2$ in spherical coordinates, expressed
as a differential operator, is exactly the Laplace--Beltrami operator. 

In general, a Laplace--Beltrami operator can be split into a radial
part $\Delta_B'$ and a transversal part. The radial part acts on
geodesics orthogonal to some submanifold $S$, typically a sphere
centered at the origin \cite{Helgason2}.

{\bf Example:} For the usual Laplace--Beltrami operator in ${\bf
R}^3$ expressed in spherical coordinates,

\beq
\label{eq:Ex1}
\Delta_B =\partial_r^2+2r^{-1}\partial_r
+r^{-2}\left(\partial_\theta^2+{\rm cot}\theta\, \partial_\theta+
{\rm sin}^{-2}\theta \, \partial_\phi^2 \right)
\eeq
  
the first two terms 

\beq
\label{eq:radR3}
\Delta_B'=\partial_r^2+2r^{-1}\partial_r
\eeq

constitute the radial part with respect to a sphere centered at the
origin and the expression in parenthesis multiplied by $r^{-2}$ is the
transversal part. The transversal part is equal to the projection of
$\Delta_B $ on the sphere of radius $r$ and equals the
Laplace--Beltrami operator on the sphere, given for $r=1$ in
eq.~(\ref{eq:Delta_on_sphere}).  This is a general result. For any
Riemannian manifold $V$ and an arbitrary submanifold $S$, the
projection on $S$ of the Laplace--Beltrami operator on $V$ is the
Laplace--Beltrami operator on $S$ (see Helgason \cite{Helgason2},
Ch.~II, paragraph 3).

The radial part of the Laplace--Beltrami operator on a symmetric space
has the general form

\beq
\label{eq:DeltaB'} 
\Delta_B'= \frac{1}{J^{(j)}}\sum_{\alpha =1}^{r'}\frac{\partial }
{\partial q^\alpha }J^{(j)}\frac{\partial }{\partial q^\alpha }\ \ \ \ \ \ \ 
(j=0,-,+)
\eeq

where $r'$ is the dimension of the maximal abelian subalgebra ${\bf
  H_0'}$ in the tangent space ${\bf P}$ (the rank of the symmetric
space) and $J^{(j)}$ is the Jacobian given in equation~(\ref{eq:J_j})
below.  The sum goes over the labels of the independent radial
coordinates defined in subsection \ref{sec-radial}: $q={\rm
  log}h(x)=(q^1,...,q^{r'})$ where $h(x)$ is the collective radial
coordinate.  These are canonical coordinates on ${\bf H_0'}$ denoted
$(q,\alpha)\equiv {\bf q\cdot \alpha }$ in \cite{OlshPere} (see below
and in the footnote referring to equation (\ref{eq:fund})). The
adjoint representation of a general element $H$ in the maximal abelian
subalgebra ${\bf H_0'}$ follows from a form similar to
eq.~(\ref{eq:adjrad}) (with or without a factor of $i$ depending on
whether we have a compact or non--compact space), where the roots are
in the restricted root lattice.  For a non--compact space of type
$P^*$

\beq
{\rm log}h=H={\bf q\cdot H}=\left(\begin{array}{cccccc}
0 & & & & & \\ & \ddots & & & & \\ & & 0 & & & \\ & & & {\bf q\cdot \alpha } & & \\
 & & & & \ddots & \\ & & & & & -{\bf q\cdot \eta }  \end{array}\right)=
\left(\begin{array}{cccccc}
0 & & & & & \\ & \ddots & & & & \\ & & 0 & & & \\ & & & q^{\alpha } & & \\
 & & & & \ddots & \\ & & & & & q^{-\eta }  \end{array}\right)
\eeq

Hence $q^\alpha ={\bf q\cdot \alpha }$ and 

\beq
h={\rm e}^{H}=\left(\begin{array}{cccccc}
1 & & & & & \\ & \ddots & & & & \\ & & 1 & & & \\ & & & {\rm e}^{q^\alpha } & & \\
 & & & & \ddots & \\ & & & & & {\rm e}^{q^{-\eta }}  \end{array}\right)
\eeq

{\bf Example:} For the simple rank--1 algebra corresponding to the
compact group $SU(2)$, the above formulas take the form 
(cf. eq.~(\ref{eq:SU2adjoint}))

\beq
H=\theta H_1=\theta \left(\begin{array}{ccc}
0 & &  \\  & 1 & \\ & & -1 \end{array}\right),
\ \ \ \ \ \ \ 
h={\rm e}^{i\theta H_1}=\left(\begin{array}{ccc}
1 & &  \\  & {\rm e}^{i\theta } & \\ & & {\rm e}^{-i\theta }\end{array}\right)
\eeq

The radial coordinate is $q=(q^1)=\theta $.

There is a general theory for the radial parts of Laplace--Beltrami
operators \cite{Helgason2}. It is of interest to consider the radial
part of the Laplace--Beltrami operator on a manifold $V$ with respect
to a submanifold $W$ of $V$ that is transversal to the orbit of an
element $w\in W$ under the action of a subgroup of the isometry
group of $V$. Of special interest to us is the case in which the
manifold is a symmetric space $G/K$ and the Lie subgroup is $K$.

The Jacobian $J^{(j)}=\sqrt{|g|}$ (where $g$ is the metric tensor at
an arbitrary point of the symmetric space) of the transformation to
radial coordinates takes the form

\beq
\label{eq:J_j}
\begin{array}{l}
J^{(0)}(q)=\prod_{\alpha \in R^+} (q^\alpha )^{m_\alpha }\\
\\
J^{(-)}(q)=\prod_{\alpha \in R^+} ({\rm sinh}(q^\alpha ))^{m_\alpha }\\ 
\\
J^{(+)}(q)=\prod_{\alpha \in R^+} ({\rm sin}(q^\alpha ))^{m_\alpha }\end{array}
\eeq

for the various types of symmetric spaces with zero, negative and
positive curvature, respectively (see \cite{Helgason2}, Ch.~I,
par.~5).  In these equations the products denoted $\prod_{\alpha \in
R^+}$ are over all the positive roots of the restricted root lattice
and $m_\alpha $ is the multiplicity of the root $\alpha $. The
multiplicities $m_\alpha $ were listed in Table~1.  

{\bf Example:} On the hyperboloid $H^2$ with metric

\beq
g_{ij}=\left(\begin{array}{cc} 1 & 0 \\
                       0 & {\rm sinh}^2\theta \end{array}\right),\ \ \ \ \ \ \  
g^{ij}=\left(\begin{array}{cc} 1 & 0 \\
                       0 & {\rm sinh}^{-2}\theta \end{array}\right) 
\eeq

equations (\ref{eq:DeltaB'},\ref{eq:J_j}) give the radial part of
$\Delta_B$ for $H^2$. The radial coordinate is $\theta $ so we get, in
agreement with (\ref{eq:J_j})

\beq
\begin{array}{c}
J^{(-)}=\sqrt{|g|}={\rm sinh}\theta ,\ \ \ \ 
\Delta_B'={1\over {{\rm sinh}\theta }}\, \partial_\theta \, 
{\rm sinh}\theta  \, \partial_\theta 
=(\partial_\theta^2+{\rm coth}\theta \, \partial_\theta )
\end{array}
\eeq

In the same way we can also easily derive the equations
(\ref{eq:radR3}, \ref{eq:Ex2}) using (\ref{eq:DeltaB'},\ref{eq:J_j}). 
In particular, we immediately get the radial part of the
Laplace--Beltrami operator acting on the two--sphere $S^2$
transversally to a one--sphere $S^1$ around the north pole:

\beq
\label{eq:Ex2}
\Delta_B'=(\partial_\theta^2+{\rm cot}\theta\, \partial_\theta )  
\eeq

\subsection{Zonal spherical functions}
\label{sec-zonal}

The properties of the so called zonal spherical functions are
important for the research results to be discussed in
\cite{CasMag}.  Since there is
a natural mapping from the Hamiltonians of integrable
Calogero--Sutherland systems onto the Laplace--Beltrami operators of
the underlying symmetric spaces, these eigenfunctions play an
important role also in the physics of the integrable systems.
Regarding the DMPK equation for a quantum wire, the known asymptotic
expressions for these eigenfunctions allows one to solve this equation
in general or in the asymptotic regime, because of the simple
mapping from the DMPK evolution operator to the radial part of the
Laplace--Beltrami operator.  For an example of their use see
\cite{MCDMPK,MCdis,Ol}.

It is known that when $\rho $ is an irreducible representation of an
algebra, then the Casimir operator $C_{k,\rho} $ is a multiple of the
identity operator \cite{SattW,Hermann} (Schur's lemma). This means
that it has eigenvalues and eigenfunctions. Since the Casimir
operators (and consequently the Laplace operators) form a commutative
algebra, they have common eigenfunctions. There exists an extensive
theory regarding invariant differential operators and their
eigenfunctions \cite{Helgason2}.  Of particular interest are the
differential operators on a group $G$ or on a symmetric space $G/K$
that are left--invariant under the group $G$ and right--invariant
under a maximal compact subgroup $K$.  Suppose the smooth
complex--valued function $\phi_\lambda (x)$ is an eigenfunction of
such an invariant differential operator $D$ on the symmetric space
$G/K$:

\beq
D \phi_\lambda (x)=\gamma_D(\lambda ) \phi_\lambda (x)
\eeq

Here the eigenfunction is labelled by the parameter $\lambda $ and
$\gamma_D(\lambda )$ is the eigenvalue.  If in addition $\phi_\lambda
(kxk') =\phi_\lambda (x)$ ($x\in G/K$, $k\in K$) and $\phi_\lambda (e)=1$
($e=$identity element), the function $\phi_\lambda$ is called {\it
spherical}. A spherical function satisfies \cite{Helgason2}

\beq
\label{eq:integralformula}
\int_K \phi_\lambda (xky)\, dk =\phi_\lambda (x)\phi_\lambda (y)
\eeq

where $dk$ is the normalized Haar measure on the subgroup $K$.  We
will see examples of this formula below.

The common eigenfunctions of the Laplace operators are invariant under
the subgroup $K$. They are termed {\it zonal spherical
functions}. Because of the bi--invariance under $K$, these functions
depend only on the radial coordinates $h$:

\beq
\phi_\lambda (x)=\phi_\lambda (h)
\eeq

{\bf Example:} Let's study for a moment the eigenfunctions of the
Laplace operator on $G/K=SO(3)/SO(2)$.  We know from quantum mechanics
that the eigenfunctions of ${\bf L}^2$ are the associated Legendre
polynomials $P_l({\rm cos}\theta )$, and $-l(l+1)$ is the eigenvalue
under ${\bf L}^2$ (our definition of ${\bf L}$ differs by a factor of $i$
from the definition common in quantum mechanics):

\beq
\label{eq:eigenPl}
{\bf L}^2P_l({\rm cos}\theta )=-l(l+1)P_l({\rm cos}\theta )
\eeq

where ${\rm cos}\theta $ is the $z$--coordinate of the point
$P=(x,y,z)$ on the sphere of radius $1$.  In spherical coordinates,
$P=({\rm sin}\theta \, {\rm cos}\phi , {\rm sin}\theta \, {\rm
  sin}\phi , {\rm cos}\theta )$.  As we can see, the eigenfunctions
are functions of the radial coordinate $\theta $ only. The subgroup
that keeps the north pole fixed is $K=SO(2)$ and its algebra contains
the operator $L_z=\partial_\phi $. Indeed, $P_l({\rm cos}\theta )$ is
unchanged if the point $P$ is rotated around the $z$--axis.

In terms of Euler angles, a general $SO(3)$--rotation takes the form

\beq
R(\alpha, \beta ,\gamma )=g(\alpha )k(\beta )h(\gamma )=
\left(\begin{array}{ccc} {\rm cos}\alpha & 0 & -{\rm sin}\alpha \\
                               0        & 1 & 0 \\
                               {\rm sin}\alpha & 0 & {\rm cos}\alpha 
\end{array}\right)
\left(\begin{array}{ccc} {\rm cos}\beta & -{\rm sin}\beta & 0 \\
                         {\rm sin}\alpha & {\rm cos}\alpha &0 \\
                           0 & 0 & 1 
\end{array}\right)
\left(\begin{array}{ccc} {\rm cos}\gamma & 0 & -{\rm sin}\gamma \\
                               0        & 1 & 0 \\
                               {\rm sin}\gamma & 0 & {\rm cos}\gamma 
\end{array}\right)
\eeq

where $g$ and $h$ are rotations around the $y$ axis by the
angles $\alpha $ and $\gamma $ respectively, and $k$ is a rotation
around the $z$ axis by the angle $\beta $. Under such a rotation, the
north pole $(0,0,1)$ goes into $(-{\rm cos}\alpha \, {\rm cos}\beta \,
{\rm sin}\gamma - {\rm sin}\alpha \, {\rm cos}\gamma , -{\rm sin}\beta
\, {\rm sin}\gamma , -{\rm sin}\alpha \, {\rm cos}\beta \, {\rm
sin}\gamma + {\rm cos}\alpha \, {\rm cos}\gamma )$.  This means that
eq.~(\ref{eq:integralformula}) takes the form

\beq
\frac{1}{2\pi }\int_0^{2\pi }P_l(-{\rm sin}\alpha \, {\rm cos}\beta \, 
{\rm sin}\gamma + {\rm cos}\alpha \, {\rm cos}\gamma )\, d\beta = 
P_l({\rm cos}\alpha )P_l({\rm cos}\gamma )
\eeq

{\bf Example:} For the symmetric space $G/K=E_2/SO(2)$ the spherical functions
are the plane waves:

\beq
\psi (r)={\rm e}^{ikr}
\eeq
 
where $k$ is a complex number.
If $g$, $h$ denote translations in the $x$--direction by a distance
$b$, $a$ respectively, and $k$ is a rotation around the origin of
magnitude $\phi $, then the transformation $g(b)k(\phi )h(a)$ moves
the point $x\in {\bf R}^2$ by a distance $\sqrt{a^2+b^2+2ab{\rm
cos}\phi }$.  Therefore we obtain from (\ref{eq:integralformula})

\beq
\frac{1}{2\pi }\int_0^{2\pi }\psi \left(\sqrt{a^2+b^2+2ab{\rm cos}\phi }
\right) \, d\phi =\psi (a) \psi (b)
\eeq

Defining the Jacobians (\ref{eq:J_j}) as in reference \cite{OlshPere}
in terms of a parameter $a$,

\beq
\begin{array}{l}
J^{(0)}(q)=\prod_{\alpha \in R^+} (q^\alpha )^{m_\alpha }\\
\\
J^{(-)}(q)=\prod_{\alpha \in R^+} (a^{-1}{\rm sinh}(aq^\alpha ))^{m_\alpha }\\ 
\\
J^{(+)}(q)=\prod_{\alpha \in R^+} (a^{-1}{\rm sin}(aq^\alpha ))^{m_\alpha }\end{array}
\eeq

it is not hard to see that the various spherical functions listed
above are related to each other by the simple transformations

\beq
\label{eq:-to0+}
\begin{array}{l}
\phi_\lambda^{(0)} (q)={\rm lim}_{a\to 0} \phi_\lambda^{(-)}(q) \\   \\
\phi_\lambda^{(+)} (q)=\phi_\lambda^{(-)}(q)|_{a\to ia} \end{array}
\eeq   

There exist integral representations of spherical functions for the
various types of spaces $G/K$ \cite{OlshPere,Helgason2}. We will list
an integral representation only of $\phi_\lambda ^{(-)}(q)$ below,
recalling that formulas for the other types of spherical functions can
be obtained by (\ref{eq:-to0+}). If $\phi_\lambda^{(-)} (x) $ is
spherical and $h$ is the spherical radial part of $x$,

\beq
\label{eq:sphericalfunctions-}
\phi_\lambda^{(-)} (x)=\phi_\lambda^{(-)} (h)=
\int_K{\rm e}^{( i\lambda  - \rho )H}(kx)dk 
\eeq

In (\ref{eq:sphericalfunctions-}) $\lambda $ is a complex--valued
linear function on the maximal abelian subalgebra ${\bf H_0'}$ of
$i{\bf P}$ and $\rho $ is the function defined by

\beq
\label{eq:rho}
\rho=\frac{1}{2} \sum_{\alpha \in R^+}m_\alpha \alpha
\eeq

In eq.~(\ref{eq:sphericalfunctions-}) they act on the unique element
$H(kx)\in {\bf H_0'}$ such that $kx=n{\rm e}^{H(kx)}k'$ in the Iwasawa
decomposition.  It was shown by Harish--Chandra \cite{HC} that two
functions $\phi_\lambda^{(-)} (x)$ and $\phi_\nu^{(-)} (x)$ are
identical if and only if $\lambda =s\nu $, where $s$ denotes a Weyl
reflection. The Weyl group is the group of reflections in hyperplanes
orthogonal to the roots and was defined in subsection
\ref{sec-rootsp}, eq.~(\ref{eq:Weyl}).

The eigenvalues of the radial part of the Laplace--Beltrami operator
corresponding to the eigenfunctions on zero, negative and positive
curvature symmetric spaces are given by the following equations (see
\cite{OlshPere} and \cite{Helgason2}, Ch.~IV, par.~5):

\beq
\label{eq:eigen}
\begin{array}{l}
\Delta_B'\phi_\lambda^{(0)}= -\lambda^2 \phi_\lambda^{(0)}\\ \\
\Delta_B'\phi_\lambda^{(-)}= (-\frac{\lambda^2}{a^2}-\rho^2)\phi_\lambda^{(-)}\\ \\
\Delta_B'\phi_\lambda^{(+)}= (-\frac{\lambda^2}{a^2}+\rho^2)\phi_\lambda^{(+)}
\end{array}
\eeq

where $\rho $ is the function in (\ref{eq:rho}). (To avoid confusion,
note that the eigenvalue $l$ in eq.~(\ref{eq:eigenPl}) is not equal to
$\lambda $. In fact, $\lambda = l+1/2$.) 

{\bf Example:} Take the symmetric space $SO(3)/SO(2)$. From Table~1 we
see that this space has $p-q=2-1=1$ short root of length 1. Then 

\beq
\rho^2 =\left(\frac{1}{2}\right)^2 \cdot 1^2 \cdot |\alpha |^2 =\frac{1}{4}
\eeq

and setting $a=1$, the eigenvalue is $-\lambda^2+1/4=-l(l+1)$.   

\subsection{The analog of Fourier transforms on symmetric spaces}
\label{sec-Fourier}

Much of the material presented in this subsection is taken from the book
by Wu--Ki Tung \cite{Wu}.

A continuous smooth ($C^\infty$) spherical function $f$ is said to be
{\it elementary} if it is an eigenfunction of any differential
operator that is invariant under left translations by $G$ and right
translations by $K$. Thus the eigenfunctions of the Laplace operators
are elementary.  The elementary spherical functions are related to
irreducible representation functions for the group $G$. The
irreducible representation functions are the matrix elements of the
group elements $g$ in the representation $\rho $. Let's clarify this
statement by an example.

{\bf Example:} The angular momentum basis for $SO(3)$ is defined by 

\beq 
\begin{array}{l}
{\bf L}^2|lm>=l(l+1)|lm> \\ 
L_3|lm>=m|lm> \\ 
L_\pm |lm>=\sqrt{l(l+1)-m(m\pm 1)}|lm> \end{array}
\eeq

where $l$ labels the representation.  The irreducible representation
functions are, in the angular momentum basis $|lm>$, the matrix
elements $D^l(R)^{m'}_{\ m}$ such that

\beq
R|lm>=|lm'> D^l(R)^{m'}_{\ m}
\eeq

where $R={\rm exp}({\bf t\cdot L})$ is a general $SO(3)$ rotation.  It
is known that for $R\in SO(3)$, if $\alpha $, $\beta $, $\gamma $ are
the Euler angles of the rotation $R=R(\alpha ,\beta ,\gamma)$, these 
matrix elements take the form

\beq 
D^l(R)^{m'}_{\ m}= D^l(\alpha ,\beta ,\gamma )^{m'}_{\ m}={\rm
 e}^{-i\alpha m'}d^l(\beta )^{m'}_{\ m}{\rm e}^{-i\gamma m},\ \ \ \ \
 \ \ d^l(\beta )^{m'}_{\ m}\equiv <lm'|{\rm e}^{-i\beta L_2}|lm> 
\eeq

The associated Legendre functions $P_l^m({\rm cos}\theta)$ and the
special functions $Y_l^m(\theta, \phi )$ called spherical harmonics
are essentially this kind of matrix elements:

\beq \begin{array}{l}
P_l^m({\rm cos}\theta)=(-1)^m\sqrt{\frac{(l+m)!}{(l-m)!}}d^l(\theta )^m_{\ 0}\\
Y_l^m(\theta, \phi )=\sqrt{\frac{2l+1}{4\pi }}\left[ D^l(\phi ,\theta ,0)^m_{\ 0}\right]^*\end{array}
\eeq

The irreducible representation functions  
$D^l(R)^{m'}_{\ m}$ satisfy orthogonality and completeness relations. In fact,
they form a complete basis in the space of square integrable functions defined
on the group manifold. This is the Peter--Weyl theorem. From here the 
corresponding theorems follow for the special functions of mathematical
physics.

{\bf Example:} For $SO(3)$ the orthonormality 
condition reads

\beq
\label{eq:completeness}
(2l+1)\int d\tau D_l^\dagger(R)^m_{\ n}D^{l'}(R)^{n'}_{\ m'}=\delta_l^{l'} 
\delta_n^{n'} \delta_{m'}^m \ \ \ \ \ \ \  D_l^\dagger(R)^m_{\ n}\equiv
[D^l(R)^n_{\ m}]^*
\eeq

where $R=R(\alpha ,\beta ,\gamma)$ is an $SO(3)$ rotation expressed in
Euler angles and $d\tau$ is the invariant group integration measure
normalized to unity, $d\tau=d\alpha d({\rm cos}\beta )d\gamma
/8\pi^2$. That the irreducible representation functions form a
complete basis for the square--integrable functions on the $SO(3)$
group manifold can be expressed as

\beq
f(R)=\sum_{lmn}f_{lm}^nD^l(R)^m_n
\eeq

where $f(R)$ is square--integrable. Using (\ref{eq:completeness}) we obtain

\beq
f_{lm}^n=(2l+1)\int d\tau D_l^\dagger(R)^n_{\ m}f(R)
\eeq

If $R(\alpha ,\beta ,\gamma)=R(\phi, \theta ,0)$ we get the special
case of the spherical hamonics on the unit sphere (setting
$\sqrt{4\pi /(2l+1)}f_{lm}^0\equiv \tilde{f}_{lm}$):

\beq
f(\theta ,\phi )=\sum_{lm}\tilde{f}_{lm}Y_{lm}(\theta ,\phi )
\eeq

\beq
\tilde{f}_{lm}=\int  f(\theta ,\phi ) Y_{lm}^*(\theta ,\phi ) d({\rm cos}\theta )d\phi
\eeq

and further, for  $R(\alpha ,\beta ,\gamma)=R(0,\theta ,0)$ we get the 
completeness relation for the associated Legendre polynomials 
$P_l({\rm cos}\theta )=Y_{l0}\sqrt{4\pi /(2l+1)}$ 

\beq
\label{eq:sum}
f(\theta )=\sum_l f_l P_l({\rm cos}\theta ) 
\eeq

\beq
f_l={(2l+1)\over 2}\int f(\theta ) P_l^*({\rm cos }\theta ) d({\rm cos}\theta )
\eeq

where $f_l=f_{l0}^0$.
These are analogous to Fourier transforms. 
In the above example, we considered a symmetric space with positive curvature. 
For a space with zero or negative curvature, we have an integral instead of a
sum in (\ref{eq:sum}):

\beq
\label{eq:int}
f(q)=\int \tilde{f}(\lambda )\phi_\lambda^{(j)}(q)d\mu (\lambda )
\eeq

\beq
\tilde{f}(\lambda )=\int f(q)\left[ \phi_\lambda^{(j)}(q)\right]^*J^{(j)}(q)dq
\eeq

where $q$ are canonical radial coordinates.  The integration measure
$d\mu (\lambda )$ was determined by Harish--Chandra to be
well--defined and proportional to $w^2|c(\lambda )|^{-2}d\lambda $
where $c(\lambda )$ is a known function whose inverse is analytic
\cite{HC} (see also \cite{OlshPere,Helgason2}) and $w$ is the order of
the Weyl group (the number of distinct Weyl reflections).
$J^{(j)}(q)dq$ is the invariant measure on the space of radial
coordinates.  In equation (\ref{eq:int}) the arbitrary
square--integrable function $f(q)$ is expressed in terms of the
complete set of basis functions $\phi_\lambda^{(j)} (q)$.

One can show \cite{Helgason2,HC} that the dimension of the space of
eigenfunctions of $\Delta_B'$ is less than or equal to $w$. It is a
remarkable fact that the eigenfunctions of the radial part of the
Laplace--Beltrami operator $\Delta_B'$ have the property of being
eigenfunctions of the radial part of {\it any} left--invariant
differential operator on the symmetric space as well
(\cite{Helgason2}, Ch.~IV).
  
The following asymptotic expression for the zonal spherical functions
on spaces of zero and negative curvature holds for large values of $|h|$
\cite{OlshPere,HC}:

\beq
\phi_\lambda (h)\sim \sum_{s\in W} c(s\lambda ){\rm e}^{(is\lambda -\rho )(H)}
\eeq

where $h={\rm e}^H$ is a spherical coordinate, $H$ is an element of
the maximal abelian subalgebra, $\lambda $ is a complex--valued linear
function on the maximal abelian subalgebra, and the function $\rho $
was defined in eq.~(\ref{eq:rho}).

\section{Discussion}

In this review we have introduced the reader to some of the most
fundamental concepts in the theory of symmetric spaces. We have tried
to keep the discussion as simple as possible without assuming any
previous familiarity of the reader with symmetric spaces. The review
should be particularly accessible to physicists. In the hope of
addressing a wider audience, we have almost completely avoided using
concepts from differential geometry, and we have presented the subject
mostly from an algebraic point of view.  In addition we have inserted a
large number of simple examples in the text, that will hopefully help
the reader visualize the ideas.

Since our aim in the forthcoming paper \cite{CasMag} will be to
introduce our readers to the application of symmetric spaces in
physical integrable systems and random matrix models, we have chosen
the background material presented here with this in mind. Therefore we
have put emphasis not only on fundamental issues but on subjects that
will be relevant in these applications as well. 

In particular we have discussed symmetric spaces of positive, zero and
negative curvature that will be relevant for matrix models of the
circular, gaussian, and transfer matrix type, respectively. In
\cite{CasMag} we will define and discuss various types of random matrix
ensembles and their applications to various physical problems, and we
will associate them to the corresponding symmetric spaces in Table~1.
For the reader unfamiliar with random matrix theories, we will mention
here that they are used for a wide range of applications in
theoretical physics. They are successfully employed in describing
universal spectral properties of systems with a large number of
degrees of freedom, where it is difficult or impossible to solve the
physical problem exactly.  In typical applications, large matrices
with given symmetry properties and randomly distributed elements
substitute a physical operator, for example a Hamiltonian, scattering
matrix, transfer matrix, or Dirac operator.  The symmetry properties
of the physical operator determine which ensemble of random matrices
to use.  Historically, large random matrices were first employed by
Wigner and Dyson to describe the energy levels in complex nuclei,
where they modelled the Hamiltonian of the system. A major area of
application is in the description of the infrared limit of gauge
theories, where an integration over an appropriate random matrix
ensemble replaces the integration over gauge field configurations in
the partition function.  Random matrix theories are also applied in
the theoretical description of mesoscopic systems, where they may be
used to model the properties of scattering and transfer matrices in
disordered wires and quantum dots.

As mentioned in the introduction, one of the applications of symmetric
spaces in the random matrix theory of quantum transport lies in the
identification of the DMPK operator with a simple transformation of
the Laplace--Beltrami operator on the symmetric space defining the
random matrix universality class.  This connection was discussed by
H\"uffmann in \cite{Huff}, and more recently it was exploited in
\cite{MCDMPK} in solving the DMPK equation.  The DMPK equation
describes the evolution of the distribution of the set of eigenvalues
of the transfer matrix with an increasing length of the quantum wire.
The identification of the matrix eigenvalues (also called radial
coordinates) with the points in the coset manifold results in the
observation that only the radial part of any operator will influence
their dynamics. To enable the reader to appreciate these issues, we
have introduced our readers to the concepts of radial coordinates,
Laplace--Beltrami operators, and discussed the properties of zonal
spherical functions.

In \cite{CasMag} we will discuss the relationship between the
restricted root lattices of symmetric spaces and integrable
Calogero--Sutherland models describing interacting many--particle
systems in one dimension. Olshanetsky and Perelomov \cite{OlshPere}
showed that the dynamics of these systems are related to free
diffusion on a symmetric space. This relationship is due to the fact
that the Hamiltonians of integrable Calogero--Sutherland models map
onto the radial parts of the Laplace--Beltrami operators of the
underlying symmetric spaces. It is a beautiful fact that the
Calogero--Sutherland model becomes exactly integrable for values of
the coupling constants determined by the restricted roots. We may call
these values the root values.  Since the DMPK equation can be mapped
onto a Schr\"odinger--like equation in imaginary time, featuring a
Calogero--Sutherland Hamiltonian with root values of the coupling
constants, this fact can be used to extract information on the DMPK
equation of a quantum wire \cite{MCDMPK,MCdis}.  This will be
described in more detail in \cite{CasMag}.  In our future publication
we will also mention various types of potentials for the
Calogero--Sutherland models and show that the Weierstrass ${\cal
  P}$--function summarizes three such potentials in various limits.
These limiting potentials correspond to particles interacting on a
circle, hyperbola, or line, respectively.  This reflects the
triplicity of the symmetric spaces in terms of their curvature.

Finally, having discussed the applications of symmetric spaces in
connection with integrable models and quantum transport problems, and
the emerging classification of random matrix ensembles, in
\cite{CasMag} we plan to indicate some possible new directions of
research, hoping to have stimulated new interest in this intriguing
but little known research field.
 
\section{Acknowledgments}

The author wishes to thank Prof. Michele Caselle of the Department of
Theoretical Physics at the University of Torino, Prof. Simon Salamon
of the Department of Mathematics at the Polytechnic University of
Torino, and Dr. Anna Fino and Dr. Sergio Console of the Department of
Mathematics, University of Torino for discussions, for reading the
manuscript and for providing helpful suggestions.


\begin{thebibliography}{99}

\bibitem{OlshPere} M. A. Olshanetsky and  A. M. Perelomov, {\bf Phys. Rep.} 94 
(1983) 313 

\bibitem{DysonSS} F. Dyson, Comm. Math. Phys. 19 (1970) 235

\bibitem{CasMag} M. Caselle and U. Magnea (work in preparation)

\bibitem{AltZ} A. Altland and M. R. Zirnbauer, {\tt cond-mat/9602137}

\bibitem{Zirn} M. Zirnbauer, J. Math. Phys. 37 (1996) 4986

\bibitem{MCclass} M. Caselle, unpublished, {\tt cond-mat/9610017}

\bibitem{TitBrou} M. Titov, P. W. Brouwer, A. Furusaki, and C. Mudry,
{\tt cond-mat/0011146}

\bibitem{Ivanov} D. A. Ivanov, {\it ``Random matrix ensembles in
    p--wave vortices''}, contribution to the proceedings of the
  workshop "Vortices in unconventional superconductors and superfluids
  -- microscopic structure and dynamics", Dresden, March 2000, {\tt
    cond-mat/0103089}

\bibitem{MCDMPK} M. Caselle, Phys. Rev. Lett. 74 (1995) 2776, 
{\tt cond-mat/9410097}

\bibitem{MCdis} M. Caselle, Nucl. Phys. B 45A (1996) 120

\bibitem{Helgason} S. Helgason, {\it Differential Geometry, Lie Groups and 
Symmetric Spaces} (Academic, New York 1978) ISBN: 0-12-338460-5

\bibitem{Gilmore} R. Gilmore, {\it Lie groups, Lie algebras, and some of 
their applications} (John Wiley \& Sons, New York 1974) ISBN: 0-471-30179-5

\bibitem{FosNigh} J. Foster and J. D. Nightingale, {\it A short course in 
general relativity} (Longman, New York 1979) ISBN: 0-582-44194-3

\bibitem{Boothby} W. M. Boothby, {\it An introduction to differentiable 
manifolds and Riemannian geometry} (Academic Press, New York 1975)

\bibitem{3w} Y. Choquet--Bruhat, C. DeWitt--Morette, and M. Dillard--Bleick,
{\it Analysis, manifolds, and physics}, Part~I (Elsevier Science, Amsterdam
1982) ISBN: 0-444-86917-7 

\bibitem{SattW} D.H. Sattinger, O.L. Weaver:
Lie Groups and Algebras with Applications to Physics,
Geometry and Mechanics (Springer-Verlag, New York 1986) ISBN: 3540962409

\bibitem{Georgi} H. Georgi, {\it Lie algebras in particle physics} (Benjamin/Cummings, Reading, Mass., 1982) ISBN: 0-8053-3153-0 

\bibitem{Hermann} R. Hermann, {\it
Lie groups for physicists} (W.A. Benjamin Inc., New York 1966) 

\bibitem{Loos} O. Loos, {\it Symmetric spaces}, vol. II (W.A. Benjamin Inc., 
New York 1969)

\bibitem{Helgason2} S. Helgason, {\it Groups and geometric analysis: Integral 
geometry, invariant differential operators, and spherical functions} 
(Academic, New York 1984) ISBN: 0-12-338301-3

\bibitem{HC} Harish--Chandra, Am. J. Math., 1958 

\bibitem{Ol} M. A. Olshanetsky, Teor. Mat. Fiz. 95, No. 2 (1993) 341

\bibitem{Wu} Wu--Ki Tung, {\it ``Group theory in physics''} (World Scientific,
Singapore 1985) ISBN: 9971-966-57-3 

\bibitem{Huff} A. H\"uffmann, J. Phys. A23 (1990) 5733

\end{thebibliography}
\end{document}